\documentclass[journal]{IEEEtran}

\pdfoutput=1 

\usepackage{graphicx}
\usepackage[dvips]{psfrag}
\usepackage[hang]{caption2}
\usepackage{pifont}
\usepackage{latexsym}
\usepackage{inputenc}
\usepackage{subfigure}
\usepackage{cite}
\usepackage{psfrag}
\usepackage{url}
\usepackage{stfloats}
\usepackage{amstext,epsfig,amssymb,amsbsy,amsmath}
\usepackage{bbm}
\usepackage{mathrsfs}
\usepackage{array}
\usepackage{eurosym}
\usepackage{pifont}
\usepackage{rotating}
\usepackage{multirow}
\usepackage{multicol}
\usepackage{color}
\usepackage{lscape}
\usepackage{longtable}
\usepackage[para]{threeparttable}
\usepackage{verbatim}
\usepackage{import}

\def\fmInterface/{Functional Mockup Interface}
\def\fmUnit/{Functional Mockup Unit}
\def\rc/{RC}

\def\ep/{En\-er\-gy\-Plus}
\def\pFactory/{Pow\-er\-Fac\-to\-ry}
\def\pt#1{%
    \def\mParm{#1}%
    \ifx\mParm\empty Ptol\-e\-my%
    \else Ptol\-e\-my~#1%
    \fi%
    }

\let\newTerm=\emph
\def\eqnRef#1{Eq.\thinspace\ref{#1}}
\def\figRef#1{Fig.\thinspace\ref{#1}}
\def\FigRef#1{Fig.~\ref{#1}}  
\def\secRef#1{Sec.\thinspace\ref{#1}}

\def\qssOrder{M}

\def\cSt{x}
\def\cStEl#1{x_{#1}}
\def\stCt{N}
\def\cStMdlElAt#1#2{%
    \widehat{x}_{#1\mspace{1mu}[#2]}%
    }
\def\cCfElAt#1#2#3{%
    x_{#2\mspace{1mu}[#3]}^{[#1]}%
    }

\def\cStDeriv{\dot{x}}

\def\quantumEl#1{\Delta Q_{#1}}

\def\qSt{q}
\def\qStMdlAt#1{%
    \widehat{q}_{[#1]}%
    }
\def\qStMdlElAt#1#2{%
    \widehat{q}_{#1\mspace{1mu}[#2]}%
    }
\def\qCfElAt#1#2#3{%
    q_{#2\mspace{1mu}[#3]}^{[#1]}%
    }

\def\inputs{u}
\def\qInputs{\mu}
\def\qInputsMdlAt#1{%
    \widehat{\mu}_{[#1]}%
    }

\def\time{t}
\def\tSevtElAt#1#2{%
    t_{#1\mspace{1mu}[#2]}^{\mathrm{s}}%
    }
\def\tQevtElAt#1#2{%
    t_{#1\mspace{1mu}[#2]}^{\mathrm{q}}%
    }
\def\tPredQevtElAt#1#2{%
    \widehat{t}_{#1\mspace{1mu}[#2]}^{\mspace{5mu}\mathrm{q}}%
    }
\def\idxQevt{\ell}  
\def\idxRcnt{\ast}  

\def\derivFcn{f}
\def\derivFcnEl#1{f_{#1}}

\def\fcnOf#1{\left(#1\right)}  
\def\fcnOfBreathe#1{\!\left\lbrace(#1)\right\rbrace\!}  
\def\absBreathe#1{%
    \begin{vmatrix}\,#1\,\end{vmatrix}%
    }
\def\abs#1{|{#1}|}

\def\ordinal#1#2{{%
    \mathcode`+="002B
    \mathcode`-="0200
    {#1}\vphantom{.}^{\rm #2}%
    }}

\begin{document}

\title{Cyber physical modeling of distributed resources for distribution system operations}

\author{Spyros Chatzivasileiadis,~\IEEEmembership{Member,~IEEE,} Marco~Bonvini, Javier~Matanza, Rongxin~Yin, Zhenhua~Liu, Thierry~Nouidui, Emre~C.~Kara, Rajiv~Parmar, David~Lorenzetti, Michael~Wetter, and~Sila~Kiliccote,~\IEEEmembership{Member,~IEEE}

\thanks{The authors are with the Lawrence Berkeley National Laboratory, California, USA.
E-mail: \{initial.lastname\}@lbl.gov}}

\maketitle

\begin{abstract}
Co-simulation platforms are necessary to study the interactions of complex systems integrated in future smart grids. The Virtual Grid Integration Laboratory (VirGIL) is a modular co-simulation platform designed to study interactions between demand response strategies, building comfort, communication networks, and power system operation. This paper presents the coupling of power systems, buildings, communications and control under a master algorithm. There are two objectives. First, to use a modular architecture for VirGIL, based on the Functional Mock-up Interface (FMI), where several different modules can be added, exchanged, and tested. Second, to use a commercial power system simulation platform, familiar to power system operators, such as DIgSILENT Powerfactory. This will help reduce the barriers to the industry for adopting such platforms, investigate and subsequently deploy demand response strategies in their daily operation. VirGIL further introduces the integration of the Quantized State System (QSS) methods for simulation in this co-simulation platform. Results on how these systems interact using a real network and consumption data are also presented.

\end{abstract}


\section{Introduction}

Moving towards ``smarter'' grids, power systems complexity increases through the embedding of communication networks, demand side management, electric vehicles, and the stochastic nature of several renewable energy sources (RES). Simulation platforms specialized in power systems can no longer handle in an adequate way the increasing interdependencies with systems such as communications, buildings, and electric vehicles. More detailed simulation tools are necessary to study the system interdependencies and determine the appropriate control strategies for optimizing power system operation. An option is to extend the existing power system simulation tools by incorporating the dynamics of such networks inside the same simulation platform. On the other hand, research in the respective fields has developed highly detailed and reliable tools, which can simulate the behavior and control of such systems. This paper adopts the co-simulation approach, where highly developed and reliable simulation tools, specialized in the respective fields, are merged in a common co-simulation platform to study the interdependencies between systems and identify appropriate control strategies. 

Although co-simulation has found a lot of applications in e.g. the automotive industry or building controls (e.g. BCVTB \cite{Wetter_BCVTB}), in power systems it is a relatively recent field which has seen some development during the last 8 years. The approach followed in this paper is to couple a commercial power system simulation platform, widely used by power system operators, with advanced modeling tools for buidings and communication networks. The goal is to determine the impact the demand response strategies have on the network and determined optimal algorithms to utilize flexible loads for power system operation. 

Two are the main objectives. First, we aim at reducing the barriers for adoption of novel demand response and other control strategies in the daily power system operation. Coupling a trusted power system simulator, with which several power system operators are familiar, with other advanced modeling tools will help towards a wider adoption of such tools. Testing and becoming familiar with the impact of different strategies on the power system will allow the wider deployment and utilization of the energy reserves ``stored'' in buildings, e.g. in the form of thermal inertia. Second, we need a modular co-simulation architecture, which will allow the easy exchange and test of different simulation modules, as well as the easy extension with e.g. electric vehicle simulators, optimization tools, hardware-in-the-loop, etc. For this reason, we use the Functional Mock-up Interface standard, which provides a standardized interface for the coupling of several different tools.

This paper describes the Virtual Grid Integration Laboratory (VirGIL), which couples a commercial power system simulator with models for buildings and communication networks.
The goal is to estimate the impact of demand response strategies on the grid, and to determine optimal algorithms for exploiting flexible loads (for example, the thermal energy stored in buildings).
To this end, we describe a modular co-simulation architecture that allows the easy exchange of different simulation models, as well as the easy extension with, e.g., electric vehicle simulators, optimization tools, hardware-in-the-loop, and so on.

VirGIL's architecture is based on the \fmInterface/, which defines a standard interface for exposing the capabilities of a simulation tool~\cite{FMI_manual}.
FMI provides for a modular structure that allows the simple exchange and testing of different co-simulation tools.
VirGIL is implemented in the \pt{II} framework, which combines continuous and discrete-event simulation~\cite{Ptolemaeus2014}.

This paper is organized as follows. Section~\ref{sec:cosimulation_review} reviews existing co-simulation methods in power systems. Section~\ref{sec:VirGIL_framework} provides the overview of the proposed co-simulation architecture in VirGIL, while Section~\ref{secFMI} describes the Functional Mock-up Interface (FMI). Sections~\ref{sec:PowerFMU},~\ref{sec:BuildingsFMU}, and~\ref{sec:CommsFMU},~\ref{sec:ControlFMU} present respectively the development of the Power Systems, Buildings, Communications, and Control Functional Mock-up Units (FMUs). Sections~\ref{sec:QSSDes} and~\ref{sec:master_algorithm} describe the simulation of the model exchange FMUs based on the QSS algorithm and the operation of the master algorithm respectively. Section~\ref{sec:use_cases} describes simulation results based on real data for the LBNL network.Finally, Section~\ref{sec:conclusions} concludes this paper and provides an outlook for future extensions of this work.

\section{Co-simulation in Power Systems}
\label{sec:cosimulation_review}

Over the last years, several cosimulation approaches for power systems have been developed and documented in the literature. One of the first documented efforts is \cite{EPOCHS_first_cosim}, which co-simulates power and communications systems. The authors advocate the use of already existing simulation tools that excel in their respective fields instead of creating new simulation platforms (``federated approach''). In that work, power systems are simulated with fixed step through PSCAD/EMTDC and PSLF while communications simulations are carried out on the discrete event simulator ns-2. The two tools are synchronized at specific ``synchronization points'' without an implementation for a rollback function, which results in accumulation of synchronization induced inacurracies over time. The authors improved this approach in \cite{Thorp_power_comms_ISGT_2011} where they use a master algorithm with a common timeline for both modules. There are no ``synchronization points'', instead both simulators evolve synchronously in time.  

Most of the co-simulation approaches for power systems combine power system with communication network simulation (examples for distribution networks are \cite{Levesque_Joos_cosim_distribution_comms}, \cite{Borghetti_SITL_HLA_Cosim_Eurocon}). Ref.~\cite{Palensky_Cosim_opensource} reports a co-simulation approach for power systems and EV charging and control, where they also use FMI for the coupling of one of the simulation tools to the master algorithm. A survey of the latest simulation tools that are used for co-simulation in power systems is reported, among others, in \cite{Palensky_cosim_challenges}. 

This work focuses on the interactions of building models for energy consumption with distribution system models for power system operation.

Among the tools used for co-simulation, Gridlab-D is probably one of the most widespread \cite{GridLabD}. It has a flexible environment, which incorporates advanced modeling techniques, efficient simulation algorithms, but most importantly provide a simulation environment not only for power systems, but also incorporating detailed load modeling, rate structure analysis, distributed generator and distribution automation.  

In this paper, a commercial power system software, DigSilent Powerfactory, is used for power systems simulation. Building a co-simulation platform incorporating Powerfactory, a tool that several utilities trust and use in their daily operation, decreases the barriers for wider adoption of co-simulation tools from the industry. Power system operators can incorporate their version of Powerfactory with the co-simulation platform to investigate in more detail the effect of demand response signals, decide and subsequently deploy the most appropriate in real-time operation. Powerfactory has the additional advantage of being capable to model both AC and DC systems. A co-simulation approaches incorporating Powerfactory has also been documented in \cite{Sven_ICT_Powerfactory}. However, this is the first time that a modular co-simulation architecture, based on FMI, is implemented for coupling Powerfactory with the rest of the simulation tools.

Besides the development of the appropriate models and controls within each simulation tool, the focus in this paper is on the development of the wrapper functions which will make the modules compatible to the FMI standard for co-simulation. FMI provides for a modular structure of the co-simulation platform which allows the simple exchange and testing of different co-simulation tools. VirGIL's master algorithm will be Ptolemy II, which can combine both continuous and discrete-event simulation. At the same novel simulation algorithms are implemented in Ptolemy II, such as QSS (Quantized-State-Simulation) which allow for higher efficiency and faster execution times.

\section{VirGIL}
\label{sec:VirGIL_framework}

VirGIL (Virtual Grid Integration Laboratory) is a modular co-simulation platform that currently couples models of power systems, buildings, communications, and has the potential to integrate other sources or sinks on the electrical grid, including electric vehicles.
The platform will facilitate developing novel control algorithms, and optimizing power systems, buildings, communications and EV charging.

\begin{figure}[!htb]
    \centering
    \includegraphics[width=0.90\linewidth]{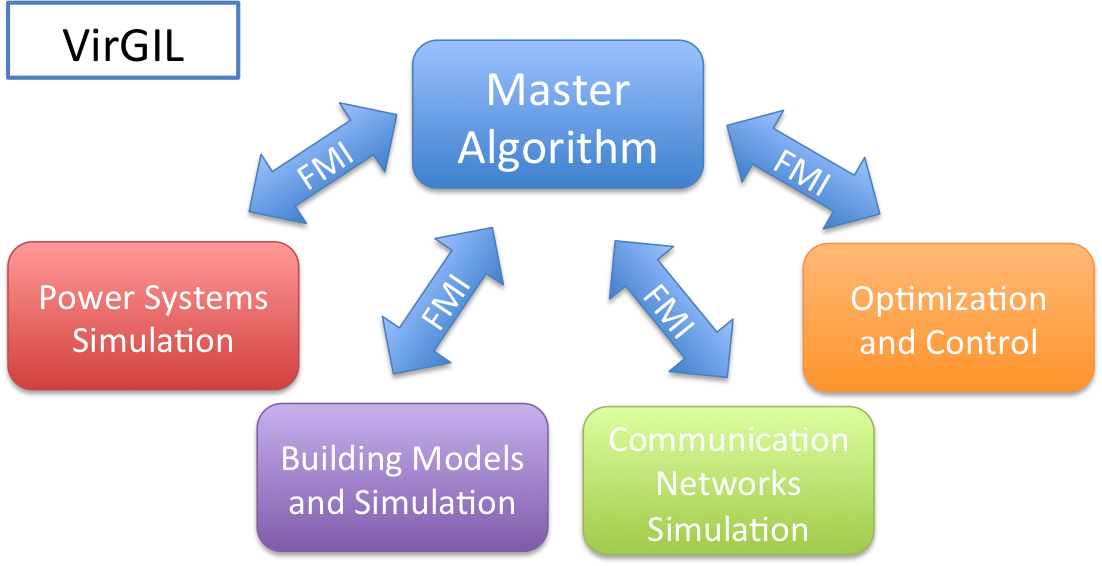}
    \caption{Overview of the VirGIL co-simulation architecture.
    Modules include tools for simulating power systems, buildings, and communications.}
    \label{fig:VirGIL_scheme}
\end{figure}

\FigRef{fig:VirGIL_scheme} shows an overview of the VirGIL co-simulation architecture.

\begin{figure}[!htb]
    \centering
    \includegraphics[width=0.90\linewidth]{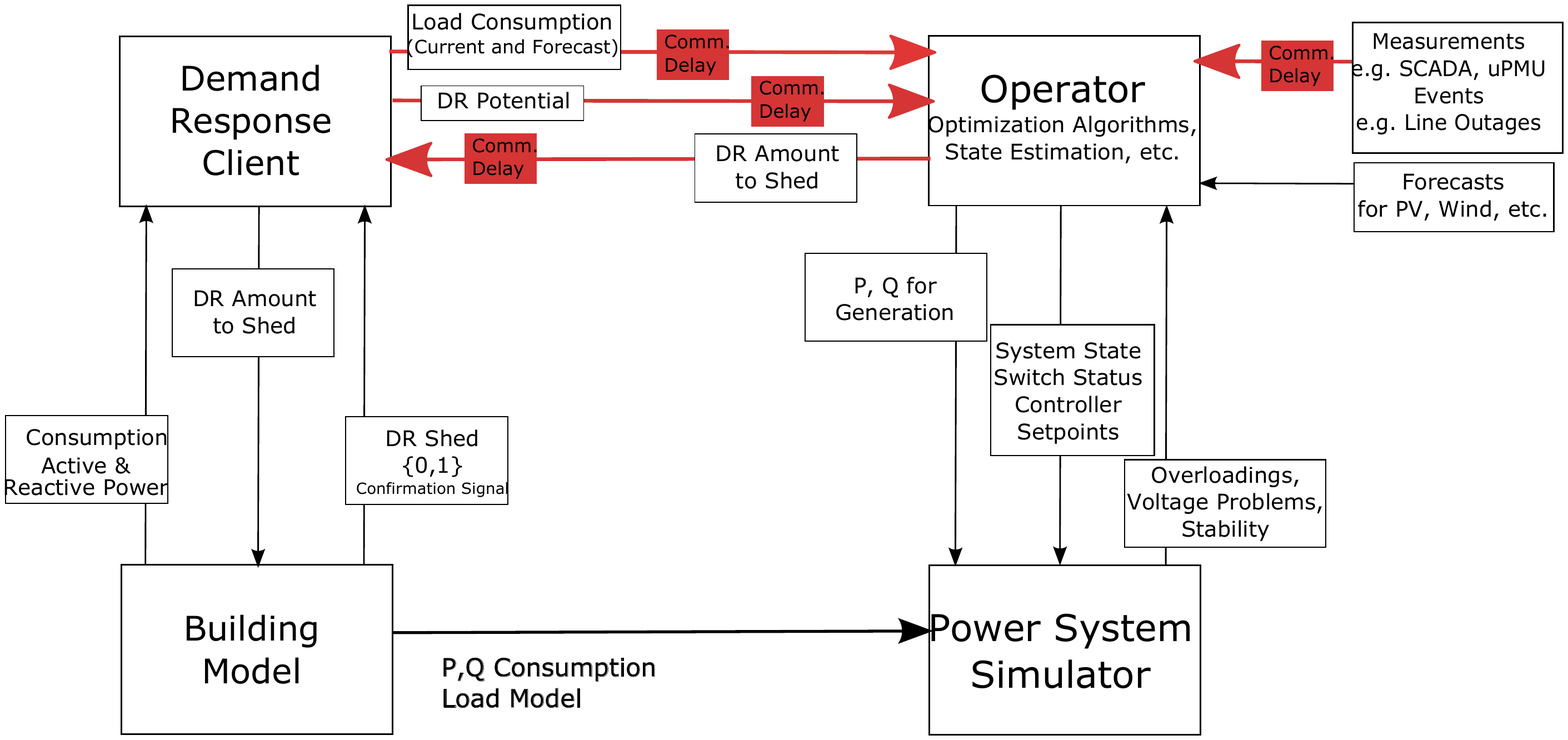}
    \caption{Interactions between VirGIL Modules}
    \label{fig:VirGIL_schema_overview}
\end{figure}

Figure~\ref{fig:VirGIL_schema_overview} presents in a schematic way the interactions between the different VirGIL blocks and the potential inputs and outputs of VirGIL. The communications modules, which is fully integrated into VirGIL, is represented in this figure as red blocks of communication delays.

\section{\fmInterface/ (FMI)}
\label{secFMI}

VirGIL's master algorithm coordinates all modules through the \fmInterface/ (FMI)~\cite{FMI_manual}.
FMI defines a tool-independent standard for exchanging models and running standalone simulation tools.
In principle, this allows VirGIL to integrate any FMI-compliant tool.
For example different power system simulation tools can be exchanged and tested without requiring changes to any other simulation module, or to the master algorithm.


A simulation model exported according to the FMI standard is called a \fmUnit/ (FMU).
An FMU is a zip file that contains the source or object code needed to execute a model, plus text files that describe the model's capabilities.
The FMU may also contain additional resources, such as documentation and auxiliary input files.
%

\begin{figure}[!htb]
    \centering
    \includegraphics[width=0.90\linewidth]{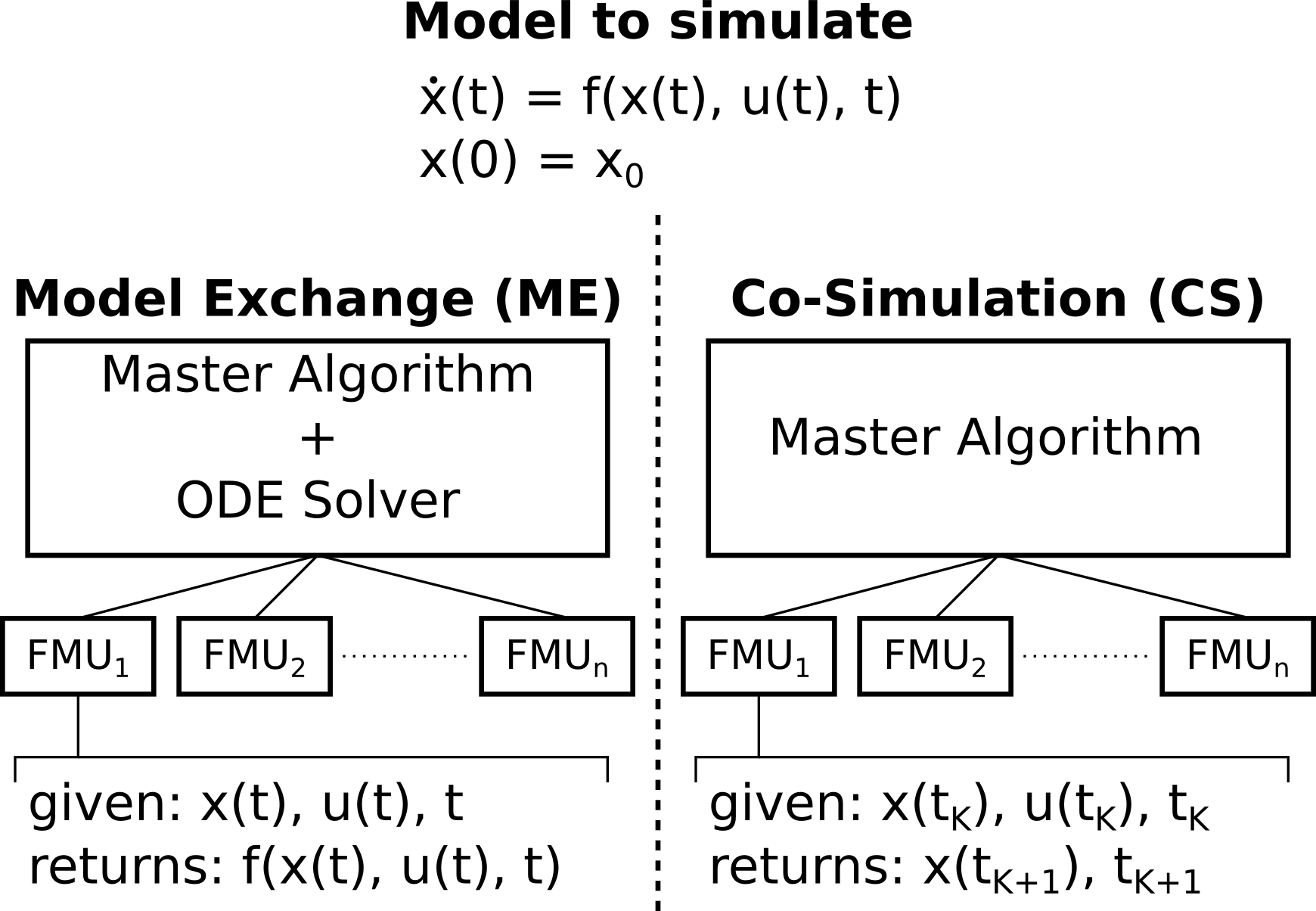}
    \caption{FMI for Model Exchange and FMI for Co-Simulation}
    \label{fig:FMI_types}
\end{figure}

The FMI standard distinguishes between Model Exchange and Co-Simulation.
See \figRef{fig:FMI_types}.
The FMI for model exchange represents a dynamic component directly, using differential, algebraic, and discrete equations.
Therefore the master algorithm must provide the necessary solvers.
By contrast, the FMI for co-simulation defines an interface for coupling independent simulation tools.
Under co-simulation, the FMU itself provides the associated solvers.
In both cases, the master algorithm coordinates time, and exchanges inputs and outputs, between FMUs.

VirGIL can integrate both types of FMU.
For example, the power system model uses the FMI for Co-Simulation, while the buildings model uses the FMI for Model Exchange.

\section{Power Systems FMU}
\label{sec:PowerFMU}

We chose DigSILENT \pFactory/ after reviewing several power system software packages.
The main focus was on established commercial power system software, in order to demonstrate how co-simulation enables the use of familiar specialized simulation tools.
For this project, \pFactory/'s scripting interfaces, for example to C++, C\#, Python, and Matlab/Simulink, made it especially attractive.
In our implementation, all VirGIL modules, except for the power systems part run on Linux. DigSILENT Powerfactory runs only on Windows. As a result, we implemented a socket communication between Windows and Linux. In Linux the PowerFMU implements all functions necessary for the FMI standard and calls their counterpart in the Windows implementation through the socket. The Windows FMU, in turn, calls the Python functions that start and stop \pFactory/, parameterize the simulation, set the inputs, and get the outputs.



\begin{figure}[!htb]
    \centering
    \includegraphics[width=0.90\linewidth]{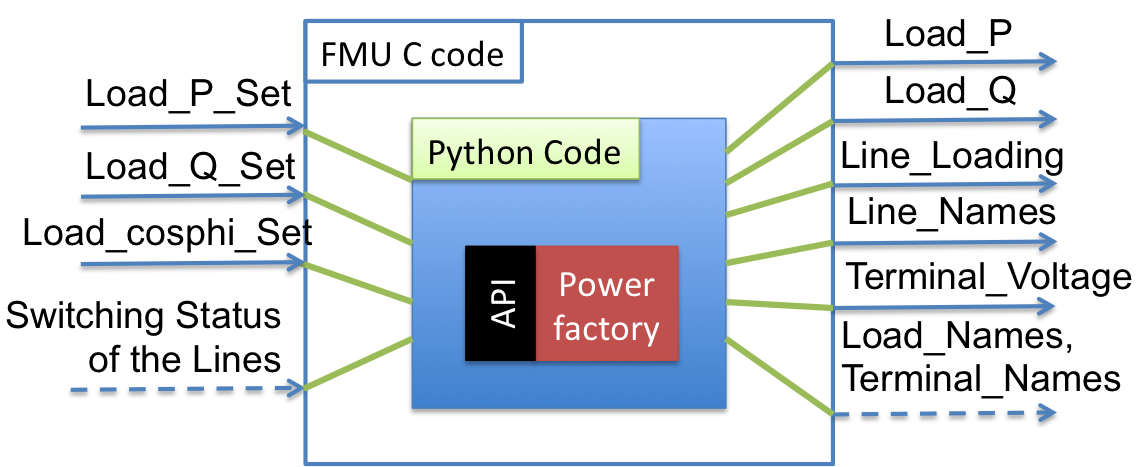}
    \caption{An FMU wrapper for \pFactory/. The arrows represent variable names.}
    \label{fig:PowerFMU}
\end{figure}

\FigRef{fig:PowerFMU} shows the structure of the FMU for power systems simulation.
The FMU maps the C-language functions defined in the FMI standard to calls on \pFactory/'s Python API.
For example:
\begin{itemize}
\item \emph{fmi2Instantiate()}: start \pFactory/, Activate Project.
\item \emph{fmi2SetReal(), fmi2SetInteger(), fmi2SetString()}: Set the values of variables and parameters.
\item \emph{fmi2DoStep()}: execute load flow.
\item \emph{fmi2GetReal(), fmi2GetInteger(), fmi2GetString()}: Get the values of variables and parameters.
\end{itemize}

VirGIL's initial focus is on the impact of demand response algorithms in the power system steady-state operation, e.g., to investigate line loadings and voltage profiles.
Thus the Power Systems FMU runs several sequential load flows, and determines the state of the system after each run.
Extending the FMU to handle dynamic simulations is an object of future work.

\section{Buildings FMU}
\label{sec:BuildingsFMU}
%

To study how demand response affects the distribution grid, VirGIL requires a 
building model that can capture the relevant dynamics, without placing undue 
computational burden on the overall simulation.
For example, the model should have sufficient detail to show the effect of DR 
strategies such as changing temperature setpoints, or reducing fan speeds.

\begin{figure}[!htb]
\centering
\includegraphics[width=0.9\columnwidth]{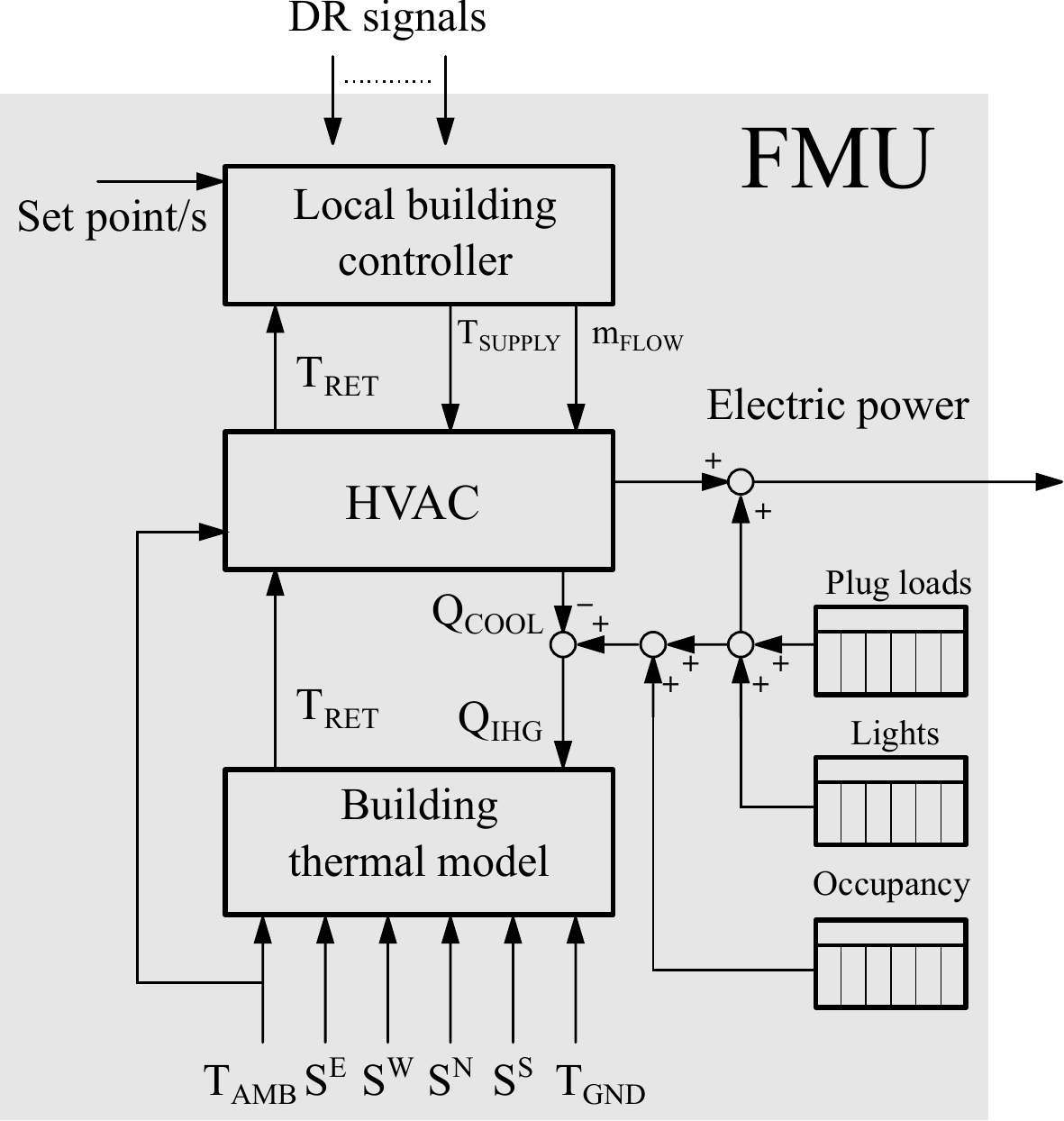}
\caption{Overview of the VirGIL building FMU model. The model comprises four main parts:
the building thermal model, the HVAC system, the schedules and the control system.}
\label{fig:building_model_overview}
\end{figure}

Building energy performance depends on the interaction between many heterogeneous 
elements, e.g., the envelope, windows, lighting, controls, and the heating 
ventilation and air-conditioning (HVAC) systems.
To represent these elements, the building model used in VirGIL comprises four 
main parts, as shown in Fig. (\ref{fig:building_model_overview}):
(1)~the thermal system that describe the envelope, windows, interior slabs and 
partitions, and room air;
(2)~the HVAC systems (e.g., air handling units, fans, etc.);
(3)~a set of schedules that describe thermal/electric loads such as lights, plug 
loads and internal heat gains generated by occupants;
(4)~and the building control systems that manages the HVAC and other assets in 
order to maintain the comfort levels and receives
DR signals.

\begin{figure}[!htb]
\centering
\includegraphics[width=2.7in]{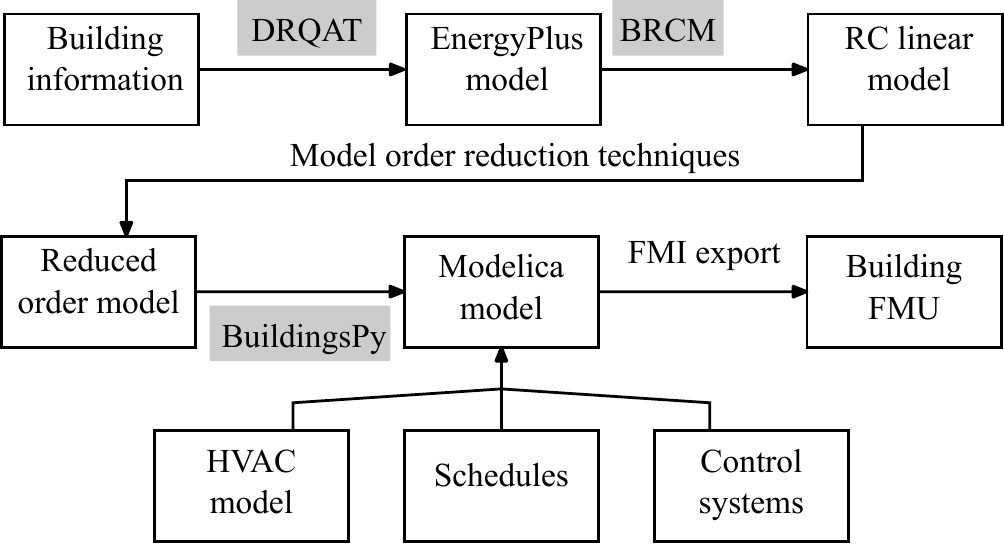}
\caption{Description of the end-to-end process for generating the building FMU 
(Grey boxes are toolboxes or packages used in the process).}
\label{fig:Building_model_creation_process}
\end{figure}

While VirGIL could incorporate \ep/ models directly, using its FMI interface for
Co-Simulation \cite{NouiduiWetter2014a}, a complete \ep/ model is too detailed 
to simulate all the buildings in a complete distribution system. The Building 
Resistance-Capacitance Modeling (BRCM) Toolbox \cite{SturzeneggerEtAl_ACC2014} 
provides an alternative to overcome the computational burden of a full-building 
simulation model such \ep/. The BRCM toolbox constitutes a part of the process 
for creating a building FMU model. Fig. (\ref{fig:Building_model_creation_process}) 
describes the end-to-end process for generating a building FMU model.
The following subsections provide a detailed description of each step of the 
process.

\subsection{Generating the \ep/ model}
An \ep/ whole-building energy simulation model is the first step towards the 
creation of a simplified building model used in VirGIL.
Creating a detailed \ep/ model can be a time consuming task, for such a reason 
the \ep/ model have been generated using the Demand Response Quick Assessment 
Tool (DRQAT)~\cite{YinEtAl2010a}. Alternately, one can use prototypical 
models~\cite{enplus_protbldgs_website}.
The generated \ep/ model contains the entire description of building geometry 
and other physical properties such as the conductivity of the wall layers, their 
thermal capacitances, the solar heat gain coefficients of the windows, etc.
These information will then be used to generate a simplified first-principle 
model of the building.

\subsection{Converting the \ep/ model to \rc/ model}
The Building Resistance-Capacitance Modeling (BRCM) toolbox allows to converts 
an \ep/ description of a building's materials and geometry, 
to a lumped-capacity \rc/ network that accounts for first-principle physical 
properties. Examples of these properties are the thermal mass and the effect 
of solar radiation.

\begin{figure}[!htb]
\centering
\includegraphics[width=0.9\columnwidth]{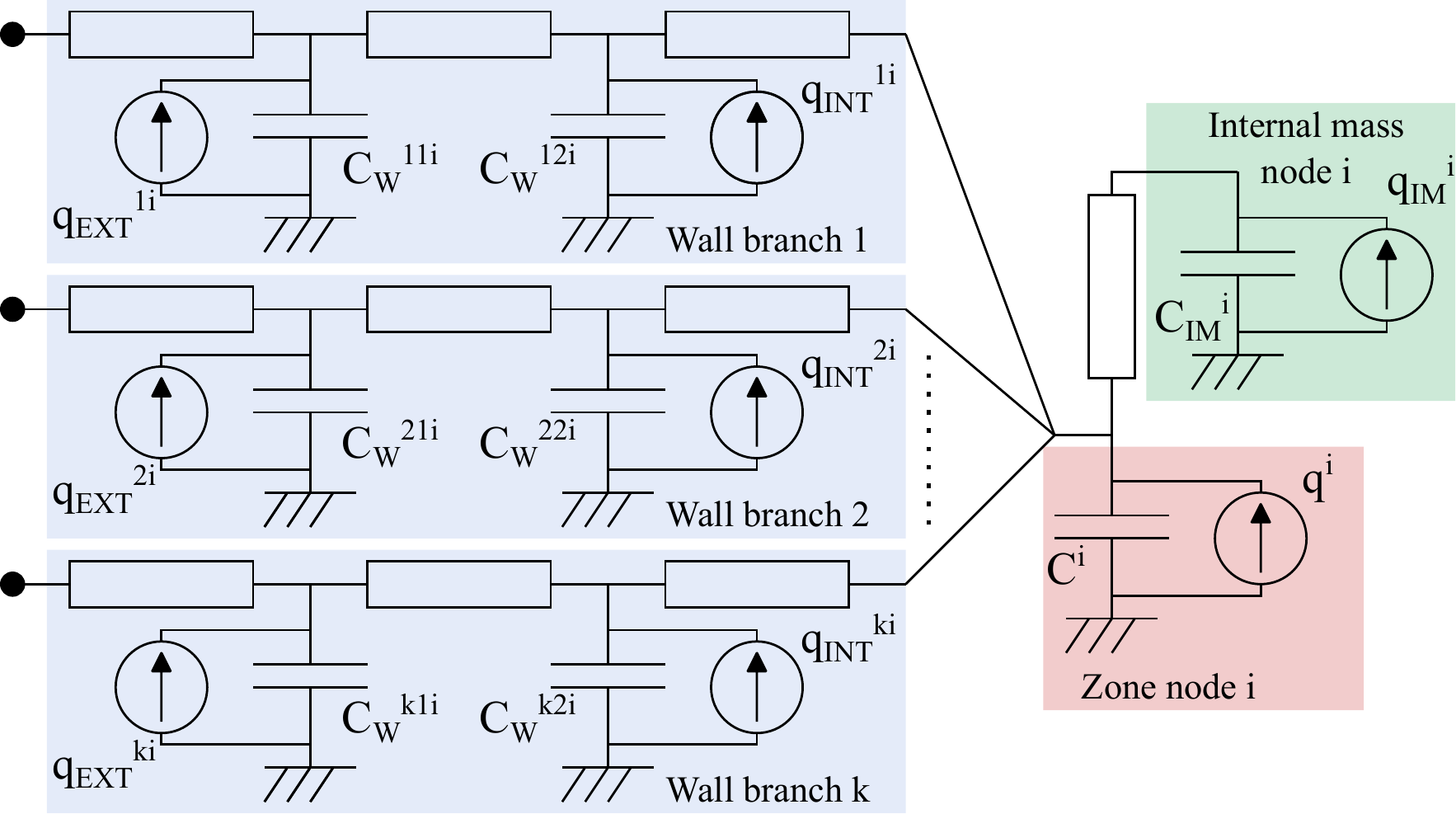}
\caption{RC network for a generic zone i. Capacitances represent states 
(i.e., Temperatures), resistances represent thermal resistances, and current 
sources represent heat fluxes.}
\label{fig:full_RC_model}
\end{figure}

For each thermal zone that is described in the \ep/ model the BRCM toolbox 
generates a RC network as shown in Fig. (\ref{fig:full_RC_model}).
For each zone $i$ the generated RC network contains the thermal capacitance of 
the air $C^{i}$, the thermal capacitance of the internal mass present in the zone
$C_{IM}^{i}$, and a series of thermal capacitances and resistances for each of 
the $N$ layers of the $k$ walls surrounding the zone $C_{W}^{kni}$.
The heat fluxes $q^{i}$,$q_{IM}^{i}$, $q_{INT}^{ki}$, and $q_{EXT}^{ki}$ 
respectively represent the internal heat gains of the zone (e.g. due to occupants, 
solar radiation, etc.), the internal heat gains heating the internal mass, 
the fraction of solar radiation directed to the innermost layer of the walls, and the
fraction of solar radiation directed the outermost layer of the walls.

Once the \rc/ network model has been parametrized the model can be written in 
the following form

\begin{subequations}
    \label{eqn:linear_full_RC}
    \begin{align}
    \dot{x}(t) &= A x(t) + B_u u(t) + B_v v(t)\\
    y(t) &= C x(t) + D_u u(t) + D_v v(t)
    \end{align}
\end{subequations}

where $x(\cdot) \in \mathbb{R}^n$ is the state vector containing all the 
temperatures of the zones, internal masses and wall layers; $u(\cdot) \in \mathbb{R}^m$ 
is the input vector (e.g., control inputs), $v(\cdot) \in \mathbb{R}^p$ are the predicted
disturbances (e.g., external air temperature, solar radiation, internal heat gains, etc.), 
and $y(\cdot) \in \mathbb{R}^o$ is the output vector.

\subsection{From \rc/ model to reduced order model}

The linear model describing the first-principle \rc/ network constitutes a first 
simplification of the whole-building model. VirGIL requires a model that is detailed 
enough to correctly capture the thermal dynamics of the buildings and correctly 
predicts the impact they have on different DR strategies. For such a purpose
the model in \eqref{eqn:linear_full_RC} needs to be further simplified.

Before starting the simplification it's important to define the outputs to be
controlled and the input control variables needed to do so. As shown in 
Fig. (\ref{fig:building_model_overview}) the building thermal model computes the
temperature of the air in the zones that is then returned to the HVAC system ($T_{RET}$).
The local controller controls the HVAC system in order to maintain the temperature
of the air in the building as close as possible to the desired set point.
The HVAC system model computes the cooling power to be delivered to maintain the 
zones temperatures at the desired set point.

This description allows to introduce two simplifications. First, the HVAC and the 
control system are not part of the building thermal model. They interact with a
suitable representation of the building that given the internal heat gains and 
the other known disturbances computes the return temperature. This allows to remove
the HVAC inputs $u(\cdot)$ from the model in \eqref{eqn:linear_full_RC}.
Second, the output vector $y({\cdot})$ is equal to the return temperature $T_{RET}$,
that is the weighted average of the  thermal zones temperature.
After introducing such simplifications the model \eqref{eqn:linear_full_RC} can be rewritten as
\begin{subequations}
    \label{eqn:linear_first_RC}
    \begin{align}
    \dot{x}(t) &= A x(t) + B_v v(t)\\
    y(t) &= C x(t) \\
    C &=
     \left(
     \begin{array}{cccccc}
       \frac{V_1}{V_{TOT}} & \cdots & \frac{V_{nx}}{V_{TOT}} & 0 & \cdots & 0
     \end{array}
    \right)
    \end{align}
\end{subequations}

where $C \in  \mathbb{R}^{1} \times \mathbb{R}^{n}$ is the output matrix, $nz$ is the number of
thermal zones (the first $nz$ elements of the state vector $x(\cdot)$), 
$V_{i}$ for $i \in [1,nz]$ is the volume of the $i$-th thermal, and 
$V_{TOT}=\sum_{i=1}^{nz}V_i$ is the sum of all the volumes. The vector of known 
disturbances $v(\cdot)$ and outputs $y(\cdot)$ are thus defined as
\begin{subequations}
    \begin{align}
    v(\cdot) &= \left(
         \begin{array}{ccccccc}
           Q_{IHG} & T_{AMB} & T_{GND} & S^{E} & S^{W} & S^{N} & S^{S}
         \end{array}
        \right)^T \nonumber \\
    y(\cdot) &= \left( T_{RET} \right) \nonumber
    \end{align}
\end{subequations}

Despite the number of input-output relationship of the model \eqref{eqn:linear_first_RC}
is seven, the number of state variables can be high enough that the simulation 
speed remains an issue (e.g., a model with ten zones can easily have more than houndred
states). For such a reason the model can be further reduced~\cite{glover1984all,zhou1993frequency}.
The resulting model will have a closely match of the input-output behaviour while 
reducing the number of states. 

\subsection{Conversion to Modelica and generation of the FMU}

Once the reduced order model that defines the input-output relationship between
the known disturbances and the output is defined, it's possible to express it
using Modelica, an object-oriented, equation-based language for modeling 
multi-domain physical systems. 

Then, drawing on the Modelica Buildings Library~\cite{wetter2014modelica}, we add 
the HVAC, loads, and controls logic.
These components predict the active power consumption of the building, and 
implement a demand response system that adjusts the zone temperatures and 
airflow setpoints according to DR signals sent by the utility.

Finally, we export the Modelica building model as an FMU for Model Exchange.

\section{Communications FMU}
\label{sec:CommsFMU}

With respect to the communication modeling, OMNeT++ \cite{OMNeT_website} was chosen among a number of simulation tools. This open-source discrete event environment is a general communication simulator widely used in the research and academic community. In this framework, a basic model is built in a hierarchical manner: first the behavior of simple modules is described in C++; then, these modules are instantiated and tied together using OMNeT++'s Network Description Language (NED) in order to form more complex entities.

Since OMNeT++'s main classes are mainly focused on the implementation of the discrete event machine and the simulation scheduler, it is common to add a number of extensions to the framework in order to upgrade the capability of the model. This is the case of the INET framework, which includes support for IPv4, IPv6, TCP, Ethernet, HTTP and many other used protocols within the Internet. Additionally, there exist other frameworks that implement mobility scenarios (like VNS), wireless sensor network (like WiXiM or Castalia), LTE technology (like SimuLTE), etc. INET counts the all the technologies that are needed for current version of VirGIL.

Other simulator options considered were: ns-2/ns-3 (Network Simulator 2 / Network Simulator 3), JiST (Java in Simulation Time) and OPNET Modeler \textregistered. Among all of them, OMNeT++ and ns-2/ns3 have extensively been used in co-simulation application for Smart Grids scenarios \cite{Godfrey2010,Muller2012,Mets2014}. There are several reasons that led us to choose OMNeT++, some of which are detailed in the following lines. 
OMNeT++ counts with a integrated development environment (IDE) adapted to Eclipse which facilitates debugging and topology creation tasks; other simulators, such as ns-2/ns-3; do not provide any kind of GUI, making debugging a very tedious task. In addition to this, OMNeT++ counts with an extensive and detailed documentation. It does not only provide information for the first steps in running a very generic simulation, but also include specific details in order to built onto the core classes for customized models. In fact, there is a specific section in the documentation on how to embed the simulation kernel into other applications, which is very helpful for implementing a co-simulation framework such as VirGIL. Once again, ns-2/ns-3 lacks organized documentation for the simulator's code. 

From a more technical perspective, OMNeT++ has shown good agreement with measured data for a number of communication technologies. This is the case of WiFi (IEEE802.11g) or LTE, as reported in \cite{Bredel2011} and \cite{Virdis2014} respectively. In terms of performance, both ns-2/ns-3 and OMNeT++ have a similar performance and offer good scalability features, as discussed in \cite{Weing2009a}.

Regarding the demand response application under study, the model counts with three high-level types of actors:  server nodes, where information about DR events is stored; client nodes, which try to retrieve this information; and a network, that interconnects all nodes. From a logical perspective, the DR communication infrastructure can be built using these three actors.

Both clients and servers in the network under study will implement Open Automated Demand Response (OpenADR) as an application layer protocol for exchanging messages. OpenADR is a standardized communications data model for sending and receiving DR signals from a utility or independent system operator to electric customers \cite{OpenADR_website,OpenADR_DRRC_website}.	

\begin{figure}[!htb]
    \centering
    \includegraphics[width=0.90\linewidth]{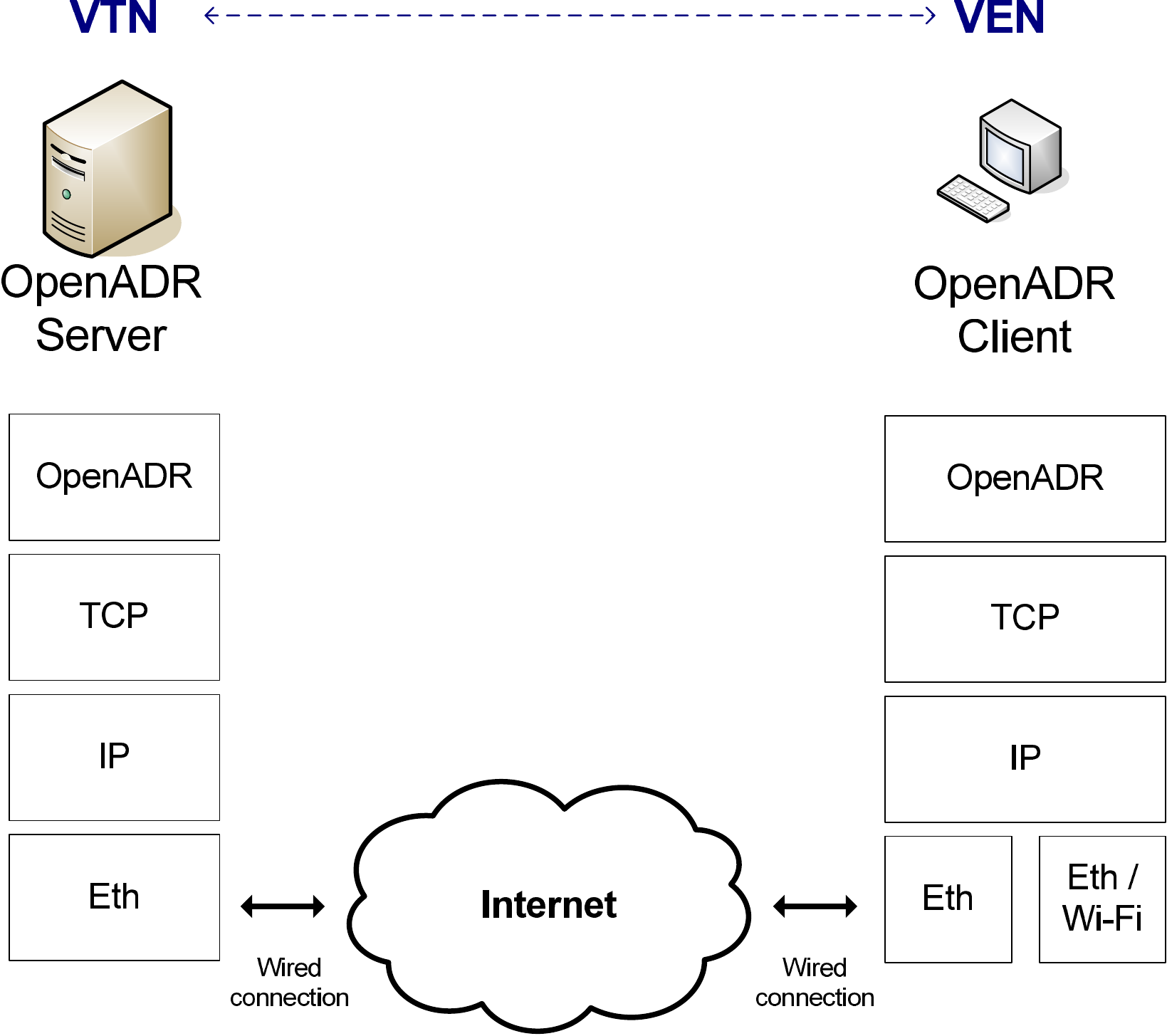}
    \caption{Communication's layer stack diagram.}
    \label{fig:networkCommStack}
\end{figure}

Figure \ref{fig:networkCommStack} shows a more detailed scheme of the model. The three already mentioned actors can be seen in the figure: an OpenADR Server, an OpenADR Client and an interconnected network (in this case the Internet). Additionally, the figure also shows the implementation of the different communication layers on each of the nodes. In this case, the {\it de facto} Internet's layer stack is chosen: TCP as a transport protocol, IP as network protocol and Ethernet as a physical protocol. OpenADR servers and clients use the lower layers to transmit their information. This layered structure, in practice produces a virtual direct communication between pairs of layers. 

Additionally, Figure \ref{fig:networkCommStack} shows the names that OpenADR' specification gives to the different nodes in the network: Virtual End Nodes (VEN) and Virtual Top Nodes (VTN). Information flows from VTN to VEN. Additionally, a VEN may also behave as a VTN in order to forward certain data to other nodes.

\begin{figure}[htb]
    \centering
    \includegraphics[width=.9\linewidth]{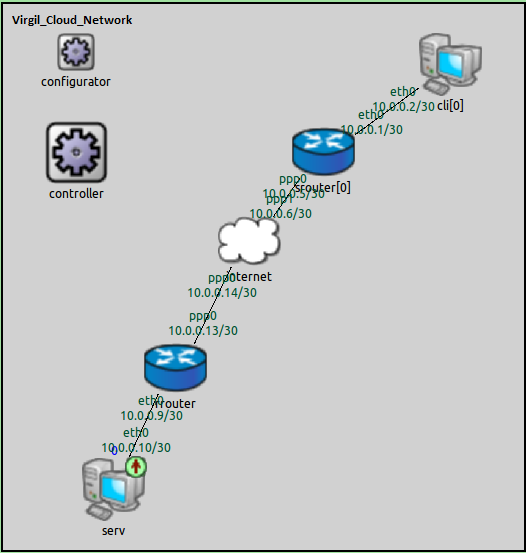}
    \caption{OMNeT++ implementation of the network.}
    \label{fig:network_OMNeT}
\end{figure}

The implementation of the network is shown in Figure \ref{fig:network_OMNeT}. It counts with a DR Server (labeled with "\texttt{serv}"), a DR Client (labeled with "\texttt{cli[0]}"), a number or routers and a cloud network. This scheme represents the communication of both nodes in an interconnected wired network such as the Internet. The figure only shows one client for clarity reasons; however, the number of clients is a parameter for the model. In case that more than one should exist, each one of them would have its own router to connect to the cloud.  

Routers pretend to simulate the gateway that each ISP (Internet Service Provider) would provide to a customer in order to connect to the Internet; as such, routers only implement up to layer-3 capabilities. The cloud network models the Internet as a network with a variable delay, transmission speed and error rate. Recalling the classification made in \cite{Mets2014} about the network model's level of detail: the Internet would be modeled as a {\it black-box} communication network, whereas the rest of the entities would count with a high level of detail (i.e. all layers and communication processes are taken into account).

\begin{figure}[tb]
    \centering
    \includegraphics[width=\linewidth]{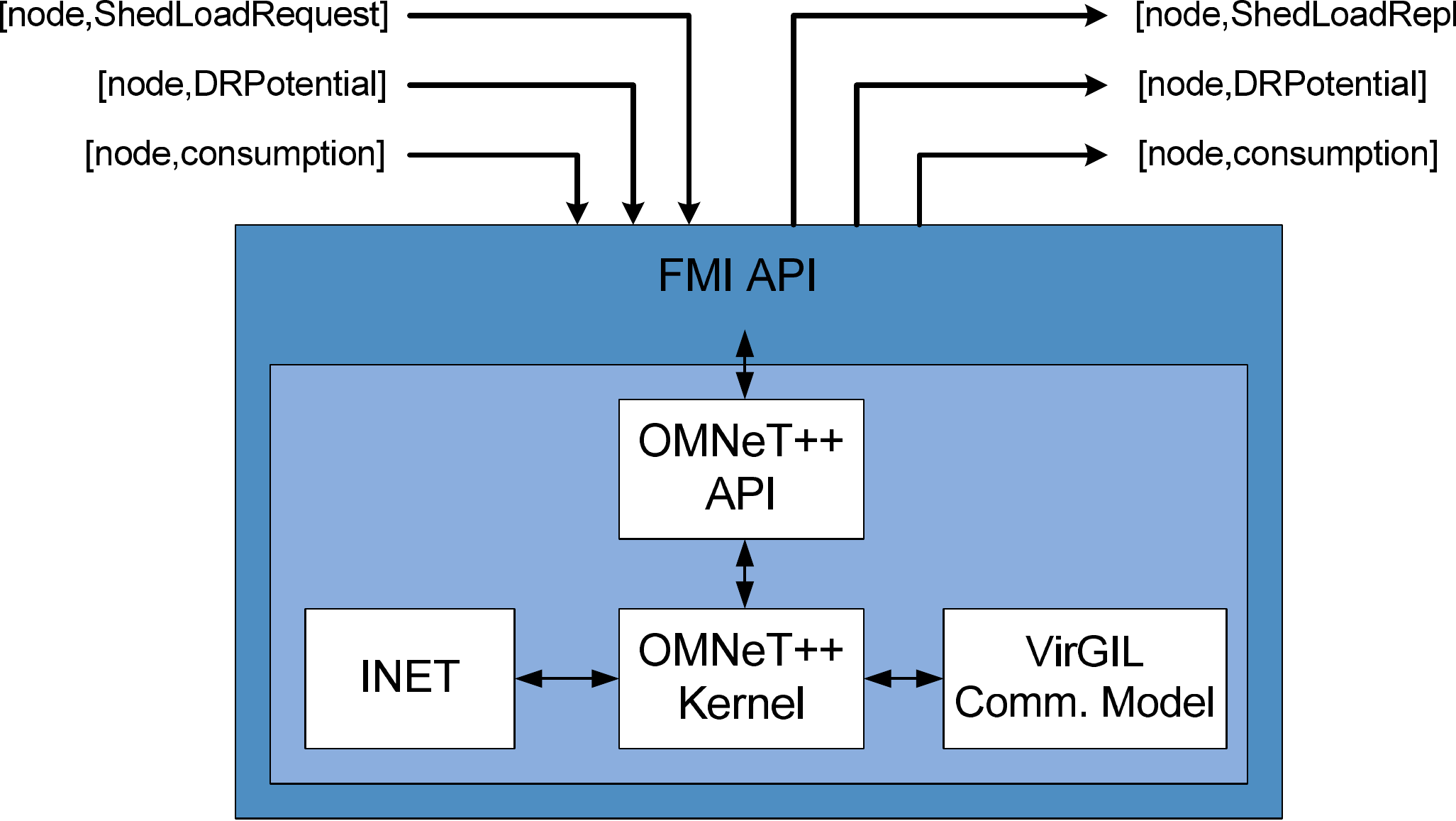}
    \caption{Structure for the Communication's FMU.}
    \label{fig:commFMU_structure}
\end{figure}

The structure of the FMU for the communications model is shown in Figure \ref{fig:commFMU_structure}. The FMU acts as an interface for the OMNeT++'s API. This API talks directly to the simulation kernel in order to set or get certain variable's values or messages. As mentioned before, the kernel implement's some functions needed in the simulation like the message scheduler or the discrete event machine. However, in order to model the Internet layer stack shown in Figure \ref{fig:networkCommStack}, the INET library is used. In addition to INET, the kernel also uses an additional library developed for this study where other capabilities (such as the OpenADR Server and Client) have been implemented. 

In order to emulate the transmission of information in the network, several parameters are input and output to and from the communication's. As seen in Figure \ref{fig:commFMU_structure}, both {\it Consumption} and {\it DRPotential} are inserted into de FMU together with a node identifier. The FMU assumes these are values transmitted from client nodes to the server node. The model simulates the transmission and, after a certain amount of time (due to the communication delay) it produces an output directed to the server. 

While {\it Consumption} and {\it DRPotential}'s information flows in one direction only, communication for DR events goes both ways. Now it is the server node which insert a {\it shedLoadRequest} message issued to a given client. Upon reception, the client decides whether to accept or not to participate in the event and replies to the server accordingly. This reply is output from the simulator after the corresponding transmission time. 

It may happen that, due to errors in the transmission, some of these messages get lost. However, these lost messages are identified by the automatic repeat request (ARQ) mechanism implemented on the TCP layer of all nodes (see Figure \ref{fig:networkCommStack}). Without going into too much detail about ARQ, whenever a message is lost, a re-transmission mechanism is trigger at the transmitting party. The result is that messages are always delivered even in the presence of errors. The only effect is that erroneous messages are affected by a higher latency (due to the re-transmission).

\section{Optimization and Control FMU} 
\label{sec:ControlFMU}

The control FMU continuously monitors the status of the integrated system and issues control signals, e.g., asking some building to increase/decrease her demand by a certain amount/percentage, charging/discharging energy storage, to ensure the health of the system.
The control signal is based on an optimization problem, e.g., optimal power flow (OPF) problem, of the power system. Generally, it is in the following form:
\begin{align}
& & \min f(P,Q,V,\theta)\\
& & k(P,Q,V,\theta) = 0\\
& & \underline{V} \leq V \leq \overline{V}\\
& & |S(\theta, V)| \leq \overline{S}
\end{align}
$f(P,Q,V,\theta)$ is the system cost, which can be the generation costs, system losses, and other control efforts. $k(P,Q,V,\theta)=0$ represents the Kirchhoff Laws. We also have network constraints, e.g., the voltage constraint and line capacity constraint, as well as other constraints not listed here. Then given system configuration and current status, the control FMU will issue control signals trying to move the system towards the optimal point of the optimization problem. Notably there are lots of challenges in solving the general form of the optimization problem. We employ the state-of-the-art technique~\cite{OPFCaltech14} to obtain the solution. It is our ongoing work to explore distributed control and decision making under uncertainty.

\begin{figure}[tb]
    \centering
    \includegraphics[width=0.8\columnwidth]{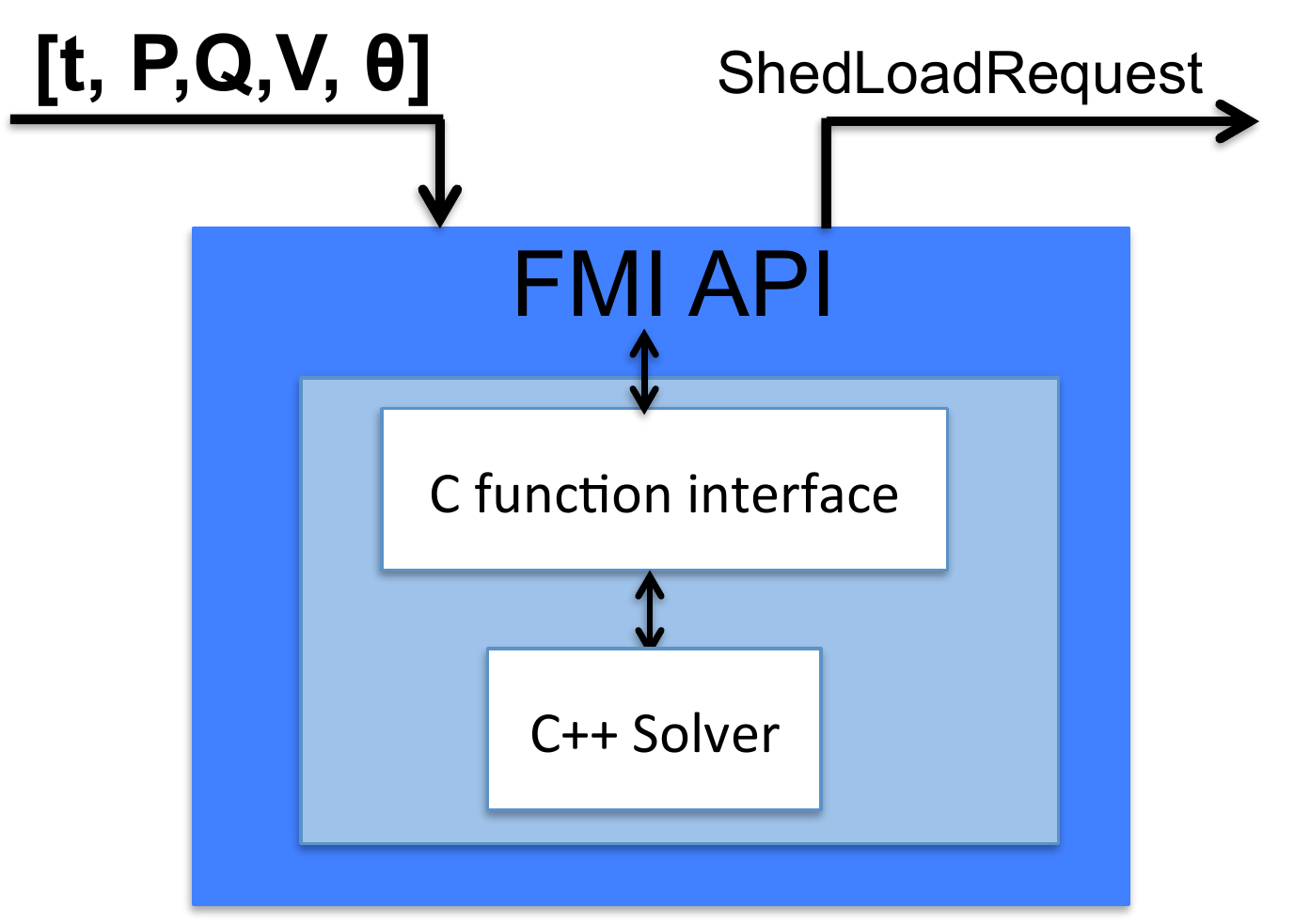}
    \caption{Structure of the Control FMU.}
    \label{fig:controlFMU_structure}
\end{figure}

The basic structure of the control FMU is shown in Figure~\ref{fig:controlFMU_structure}. The control FMU takes the system status at time $t$, e.g., $[t,V,P,Q,\theta]$, as input, and then employs the C function interface to finally call C++ solver to get the output and issue the ShedLoadRequest.

We use the following controllers for our simulations:
\begin{itemize}
\item Line capacity controller: make sure the power flow on each line is within its capacity, otherwise issue control signal to shed building load so that all line capacity constraints become satisfied. The shed request can be in either kW or percentage of current building load.
\item Volt/var controller: issue control signals to dynamically adjust the reactive power, e.g., from energy storage, to stabilize the voltage on target buses.
\item Slope controller: make sure the change slope of power consumption, voltage, and/or current is within acceptable region, otherwise issue control signal to shed building load and/or control energy storage so that the change is not too aggressive, which results in high cost due to reserves in power system.
\end{itemize}

\section{Time integration of differential equations using Quantized State System methods}
\label{sec:QSSDes}

As described in \secRef{secFMI}, each Buildings FMU defines ordinary differential equations of the form
\begin{equation}
\label{eqnDefineOde}
\cStDeriv
=
\derivFcn\fcnOf{\cSt, \inputs, \time}
\end{equation}
where
$\cStDeriv\fcnOf{\time}$ is a vector of $\stCt$ \newTerm{state variables} whose values the solver will predict;
$\inputs\fcnOf{\time}$ is a vector of \newTerm{input variables} which act as boundary conditions;
and
$\derivFcn$ is the \newTerm{derivative function}.
If \pt{} coordinates more than one such FMU, then the state variables predicted by one FMU may appear as the input variables of another.

To integrate these equations, we implemented both explicit and linearly-implicit Quantized State System (QSS) methods in \pt{}~\cite{CellierKofman2006,Migoni2013}.
QSS differs from typical integration methods, in that it discretizes the state variables rather than time.
Thus \eqnRef{eqnDefineOde} becomes
\begin{equation}
\cStDeriv
=
\derivFcn\fcnOf{\qSt, \qInputs, \time}
\end{equation}
where
$\qSt\fcnOf{\time}$ is the \newTerm{quantized state}, i.e., a discretized version of $\cSt\fcnOf{\time}$.
Likewise, $\qInputs\fcnOf{\time}$ is a quantized version of $\inputs\fcnOf{\time}$.

Quantization consists of representing a variable as a series of piecewise-continuous polynomials.
Component $j$ of the ODE system has a \newTerm{quantized state model}
\begin{equation}
\qStMdlElAt{j}{\idxQevt}\fcnOf{\time}
=
\sum_{i=0}^{\qssOrder-1} \qCfElAt{i}{j}{\idxQevt}\left(
    \time -\tQevtElAt{j}{\idxQevt}
    \right)^{i}
\end{equation}
where
$\qCfElAt{i}{j}{\idxQevt}$ denotes the $\ordinal{i}{th}$ polynomial coefficient for the $\ordinal{\idxQevt}{th}$ model;
$\tQevtElAt{j}{\idxQevt}$ gives the \newTerm{quantization-event time} at which the model was formed;
and
$\qssOrder$ gives the QSS method order.
For example, QSS1, a first-order method, quantizes the state as a constant, $\qStMdlElAt{j}{\idxQevt}\fcnOf{\time} =\qCfElAt{0}{j}{\idxQevt}$.
Each model holds on $\tQevtElAt{j}{\idxQevt} \le\time <\tQevtElAt{j}{\idxQevt+1}$ (although $\tQevtElAt{j}{\idxQevt+1}$ is not known at time $\tQevtElAt{j}{\idxQevt}$).

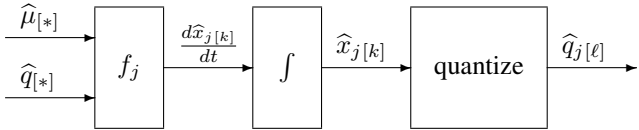
\begin{figure}[!htb]
\begin{center}
%

\setlength{\unitlength}{1cm}
\begin{picture}(8.4,1.6)

\put(0,1.2){\vector(1,0){1.2}}
\put(0.2,1.4){\parbox[b]{5em}{$\qInputsMdlAt{\idxRcnt}$}}
\put(0,0.4){\vector(1,0){1.2}}
\put(0.2,0.6){\parbox[b]{5em}{$\qStMdlAt{\idxRcnt}$}}

\put(1.2,0){\framebox(0.9,1.6){$\derivFcnEl{j}$}}

\put(2.1,0.8){\vector(1,0){1.2}}
\put(2.3,1.0){\parbox[b]{5em}{$\frac{d\cStMdlElAt{j}{k}}{d\time}$}}

\put(3.3,0){\framebox(0.9,1.6){$\int$}}

\put(4.2,0.8){\vector(1,0){1.2}}
\put(4.4,1.0){\parbox[b]{5em}{$\cStMdlElAt{j}{k}$}}

\put(5.4,0){\framebox(1.8,1.6){quantize}}

\put(7.2,0.8){\vector(1,0){1.2}}
\put(7.4,1.0){\parbox[b]{5em}{$\qStMdlElAt{j}{\idxQevt}$}}

\end{picture}
\end{center}
\caption{
QSS integration of a component of an ODE system.
The quantized state is a piecewise-continuous approximation to $\cSt\fcnOf{\time}$.
The simulation iteratively updates the state and quantized state models for individual components.
\label{figQssBlockDiagramIterative}
}
\end{figure}

Integrating the quantized state gives a series of \newTerm{state models}
\begin{equation}
\cStMdlElAt{j}{k}\fcnOf{\time}
=
\sum_{i=0}^{\qssOrder} \cCfElAt{i}{j}{k}\left(
    \time -\tSevtElAt{j}{k}
    \right)^{i}
\end{equation}
valid on $\tSevtElAt{j}{k} \le\time \le\tSevtElAt{j}{k+1}$.
Note that the \newTerm{state-event times} $\tSevtElAt{j}{k}$ may differ from the quantization-event times.
\FigRef{figQssBlockDiagramIterative} shows a block diagram of a QSS integrator.

Component $j$ forms a new state model when a quantized input to $\derivFcnEl{j}$ changes.
At the $\ordinal{k}{th}$ state-event time, the new state model is made continuous with the previous one, and its slope found from the derivative function:
\begin{align}
\cCfElAt{0}{j}{k}
&=
\cStMdlElAt{j}{k-1}\fcnOf{\tSevtElAt{j}{k}}
\\
\cCfElAt{1}{j}{k}
&=
\derivFcnEl{j}\fcnOfBreathe{\qStMdlAt{\idxRcnt}, \qInputsMdlAt{\idxRcnt}, \tSevtElAt{j}{k}}
\end{align}
where
the models $\qStMdlAt{\idxRcnt}$ and $\qInputsMdlAt{\idxRcnt}$ are evaluated at $\tSevtElAt{j}{k}$.
Index ``$\idxRcnt$'' indicates the most recent model for each component.
For QSS2 and QSS3, the higher-order coefficients $\cCfElAt{2}{j}{k}$ and $\cCfElAt{3}{j}{k}$ are estimated by perturbing the arguments to the derivative function; details are beyond the scope of this paper.

Component $j$ forms a new quantized state model when the current quantized state model differs from $\cStMdlElAt{j}{k}$ by an amount $\quantumEl{j}$, called the \newTerm{quantum}.
In the absence of other events, this happens when
\begin{equation}
\absBreathe{
    \cStMdlElAt{j}{k}\fcnOf{\tPredQevtElAt{j}{\idxQevt+1}}
    -
    \qStMdlElAt{j}{\idxQevt}\fcnOf{\tPredQevtElAt{j}{\idxQevt+1}}
    }
=
\quantumEl{j}
\end{equation}
where
$\tPredQevtElAt{j}{\idxQevt+1}$ is the \newTerm{predicted quantization-event time} for component $j$.
In practice, $\quantumEl{j}$ varies with the magnitude of $\qCfElAt{0}{j}{k}$, according to user-defined tolerances.

At each time step, the simulation advances to the minimum predicted quantization-event time from among all the components.
Thus a given global time step may re-quantize only one out of all the components.

When component $j$ does finally experience a quantization-event, it forms a new quantized state model by matching the value and derivatives from the current state model:
\begin{align}
\qCfElAt{0}{j}{\idxQevt}
&=
\cStMdlElAt{j}{k}\fcnOf{\tQevtElAt{j}{\idxQevt}}
\\
\qCfElAt{1}{j}{\idxQevt}
&=
\frac{d\cStMdlElAt{j}{k}}{d\time}\fcnOf{\tQevtElAt{j}{\idxQevt}}
\end{align}
and so on, for derivatives up to $\qssOrder-1$ (however, the linearly-implicit QSS methods offset the initial value by up to $\quantumEl{j}$).
The new quantized state model is then broadcast to any other component whose derivative function depends on $\cStEl{j}$.
This, in turn, induces state-events in those downstream components.

\begin{figure}[tb]
\begin{center}
\includegraphics[width = 0.9\columnwidth]{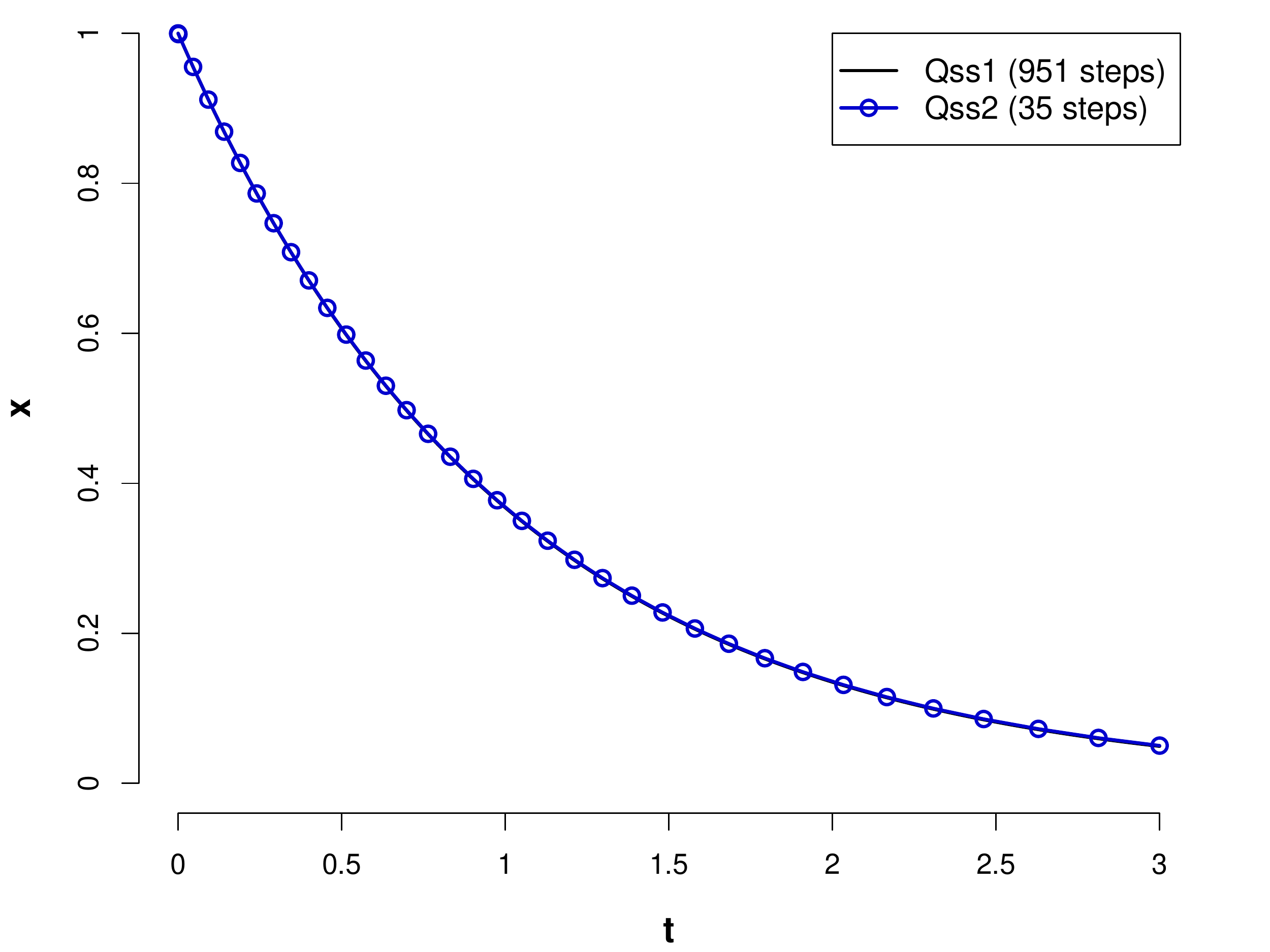}
\end{center}
\caption{QSS solution of the exponential problem, $\cStDeriv =-\cSt$.
}\label{figRunExpPr}
\end{figure}

\FigRef{figRunExpPr} shows the QSS1 and QSS2 solutions of the exponential problem, $\cStDeriv =-\cSt$, with initial condition $\cSt\fcnOf{0} =1$.
Quantum was chosen as the minimum of $0.001$ and $0.001{\cdot}\abs{\qCfElAt{0}{}{\idxQevt}}$.
Compared to the analytical result $\cSt =e^{-\time}$, both solutions end at $\time =3$ with a global error less than $5{\cdot}10^{-4}$.

The QSS approach treats every differential equation as a discrete event actor, generating events, and responding to the events produced by other equations.
However the \pt{} implementation currently groups the equations by FMU.
Thus if one equation experiences a state-event, it updates the state models for all equations contained in the same FMU.

In addition to the differential equations defined by Model Exchange FMUs, \pt{} also must handle Co-Simulation FMUs.
As suggested by \figRef{fig:PowerFMU}, the Power Systems FMU defines a static relation, determining the power flows as an algebraic function of its inputs.
Since all feedback paths from the Power System outputs back to its inputs pass through the building models, \pt{} does not have to solve any algebraic loops.
To avoid having to call the Power FMU every time a building model updates one of its outputs, we sample the building loads at discrete intervals.


\section{Master Algorithm}
\label{sec:master_algorithm}

To synchronize the data of the different FMUs,
we will use Ptolemy II~\cite{Ptolemaeus2014}.
Ptolemy II is a modular software environment 
for the design and analysis of heterogeneous systems. It provides a 
graphical model building environment, synchronizes the exchanged data and 
visualizes the system evolution during run-time. In Ptolemy II,
components are encapsulated as actors which communicate with other 
actors through ports. A director orchestrates the
data exchange between the actors and advances time for the
individual actors.\\

Next, we will discuss the mathematical structure of each FMU, and then
discuss how we componsed them for a co-simulation.
To compose multiple actors in order to conduct a co-simulation,
we need to make the distinction between
outputs of actors that directly depend on inputs, e.g.,
they have direct feedthrough,
and outputs of actors that do not directly depend on inputs. The latter are 
for example outputs of explicit time integrators that only change when
time is advanced, but not if an input is changed.

The power systems FMU implements an algebraic, time invariant system.
Therefore, the outputs of this FMU directly depend on the input values.

The building FMUs take as an input the control signal $y_{shed}$ and
produce as outputs the active and reactive
power $P$ and $Q$. Both do not directly depend on $y_{shed}$.
The building FMUs are exported using the FMI for Model-Exchange 2.0 standard.
When imported to Ptolemy II, they are combined with QSS integrator, 
as described in Section~\ref{sec:QSSDes}.
For the master algorithms, these QSS integrators can be abstracted
as actors that may schedule a time event whenever their input changes,
or whenever their state variables change by more than a tolerance.
Should the input change prior to such a scheduled event, then the actor
may replace this event with a new one that may happen at a different time.

The communications FMU lead to time delays in the signals.
They take
as inputs signal $u_j(t)$, for some $j \in \{1, \, \ldots, \, n\}$, where
$n$ is a fixed number of channels, and produce after some
time delay $\delta_j(t)$ the signal at the output. Hence, the output is
$y_j(t+\delta_j(t)) = u_j(t)$. For signal $j$, the time delay is a function
of all signals that have not yet been sent to their output, allowing to
model network congestion.
In our communication FMU, once $\delta_j(t)$ has been computed, it will not
be changed. Therefore, network congestion does not affect signals that
have already been received in the communication FMU but have not yet been
produced at its output.

The optimization and control FMU has discrete time semantics.
For a constant time step $\delta > 0$, and given measurement signal
$u(i \, \delta)$, with $i \in \{0, \, 1, \, \ldots\}$, it outputs the
control action
$y((i+1) \, \delta) = f(u(i \, \delta))$.

\begin{figure*}[tb]
\begin{center}
\includegraphics[width = 1.5\columnwidth]{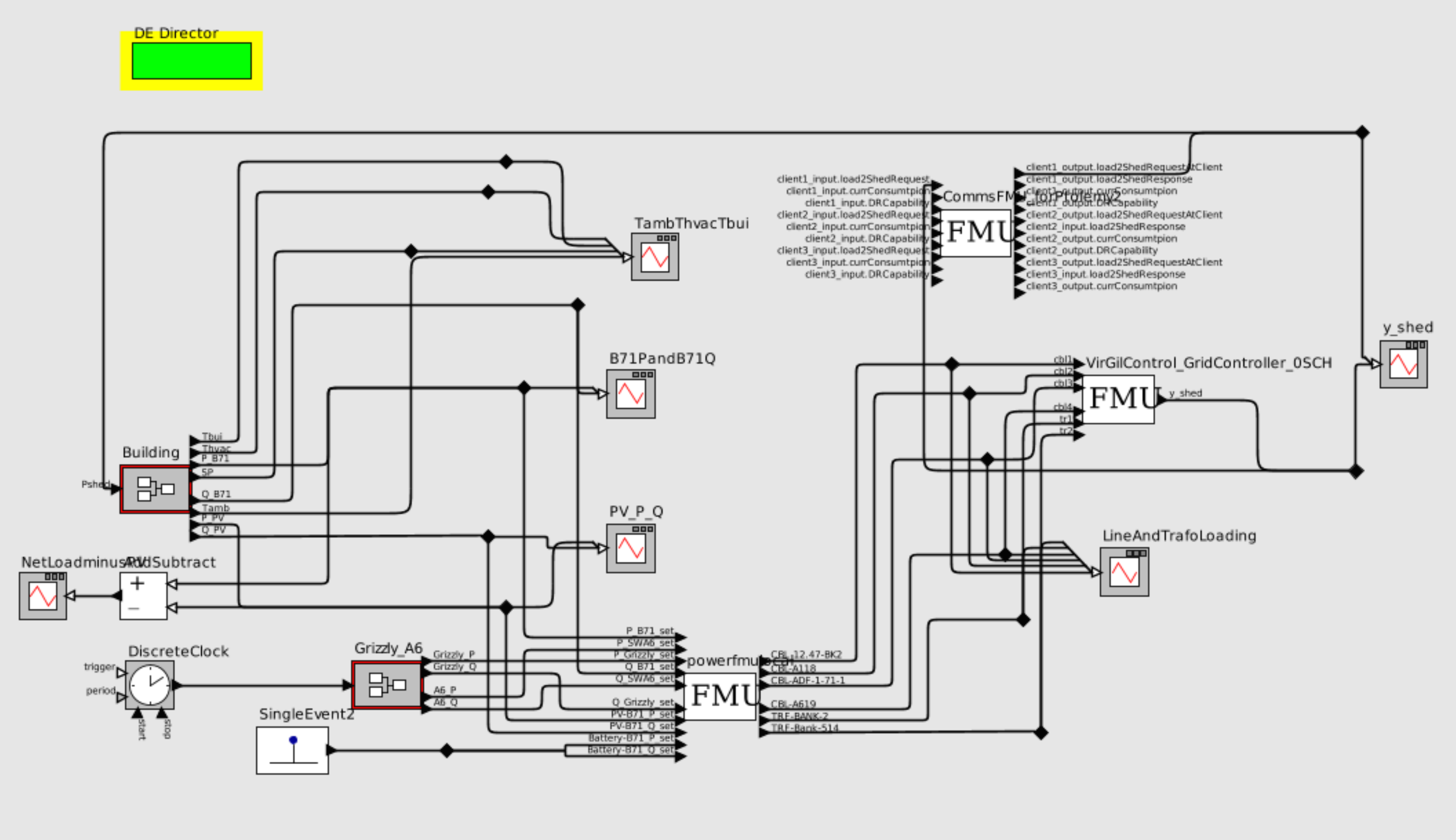}
\end{center}
\caption{Ptolemy II model that shows the composition of the different FMUs.
}
\label{fig:ptSysMod}
\end{figure*}

Figure~\ref{fig:ptSysMod} shows the Ptolemy II system model that combines
the power, building, communication and controls FMU.
The QSS Director is a new addition to Ptolemy II that we developed
in conjunction with the Ptolemy II development team.
The QSS director extends the discrete event director,
and adds a QSS solver.
Thereby, this director allows combining FMUs for model exchange,
which will be integrated with the QSS algorithm, with FMUs for co-simulation.
In addition, other Ptolemy II actors that work in the discrete event
domain can be used in such system models.

\section{Simulation Results}\label{sec:use_cases}
\subsection{Calibration of the building model}
\label{sec:building_calibration}

The case study LBNL's building~71 is a 54,000 ft\^2 two story steel-frame office and laboratory building located in Berkeley, California. The building has a water-cooled chiller system with three cooling towers. The building's operation is typical of office and laboratory, with an operational schedule of 9:00~am to 6:00~pm and high equipment usage during off-hours. The peak electric power demand is over 400 kW during the period of 12:00~pm to 6:00~pm and the average demand during the off hours is about 80 kW. 

The building FMU has been generated as described in section \secRef{sec:Buildings_FMU}. 
Before using the building FMU model we performed a calibration of the model in order to test the ability of the simplified
RC model to replicate the results of the more detailed \ep/ model. The main difference between the \ep/ model and the simplified
RC mdoel are the nonlinear relationships and complex algorithm used by \ep/ to compute interior and exterior convective 
coefficients, solar heat gain coefficients and long wave radiation effects. However, the RC model has a number of coefficients
that can be tuned in order to align the simulation results as much as possible.
Sensitivity analysis was conducted to rank the key parameters of the RC model that could be tuned. The parameters with higher sensitivity that were used to tune the model are: exterior wall convective coefficient, building solar absorption factor, window heat gain factor and heat transmission value. We therefore used the summer period from May to October to perform a comparison between the RC and the \ep/ models. We decided to use such a period because it is of interest for the demand response events as well for critical peak pricing.

Both the RC and the \ep/ model have been simulated using standard weather for San Francisco, lighting, plug loads, occupancy and set point schedules for the zones of the building. Given the same boundary condition and operation the models predicted the
cooling load required to satisfy the required comfort conditions.

\begin{figure}[!htb]
\centering
\includegraphics[width=0.9\columnwidth]{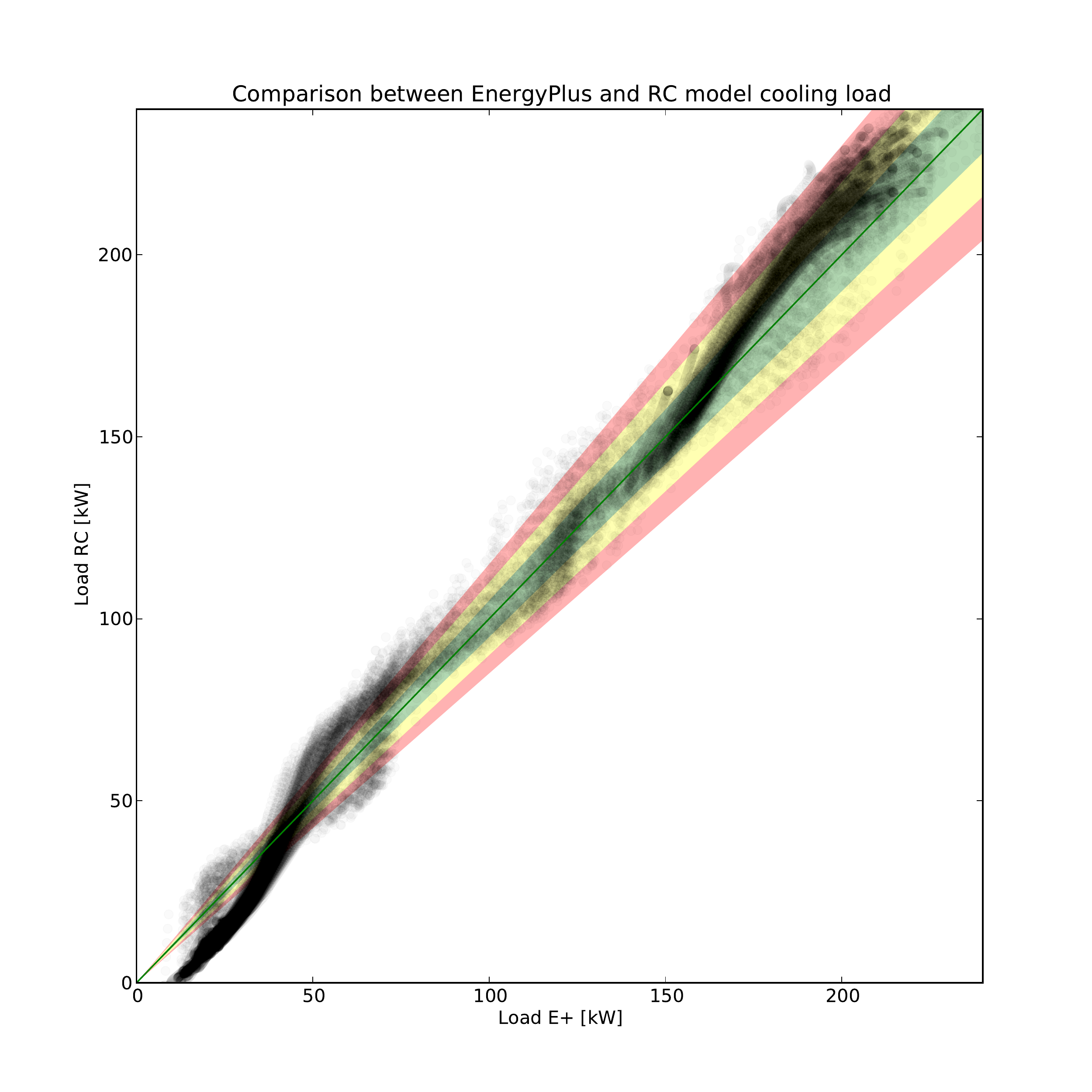}
\caption{Comparison of the cooling load predicted by \ep/ and the RC model.}
\label{fig:building_scatterplot_comp}
\end{figure}

Figure (\ref{fig:building_scatterplot_comp}) shows the RC model cooling load versus the EnergyPlus model cooling load. Every
points represents a simulated data point with a 5 min resolution over the period between May and October. The green, yellow
and red area respectively represent a relative error of $\pm5$\%, $\pm10$\% and $\pm15$\%. As can be seen the highest
relative error occurs at low cooling load level while when the load is close to its maximum the almost totality of teh points is
within the $\pm15$\% interval.

\begin{figure}[!htb]
\centering
\includegraphics[width=0.9\columnwidth]{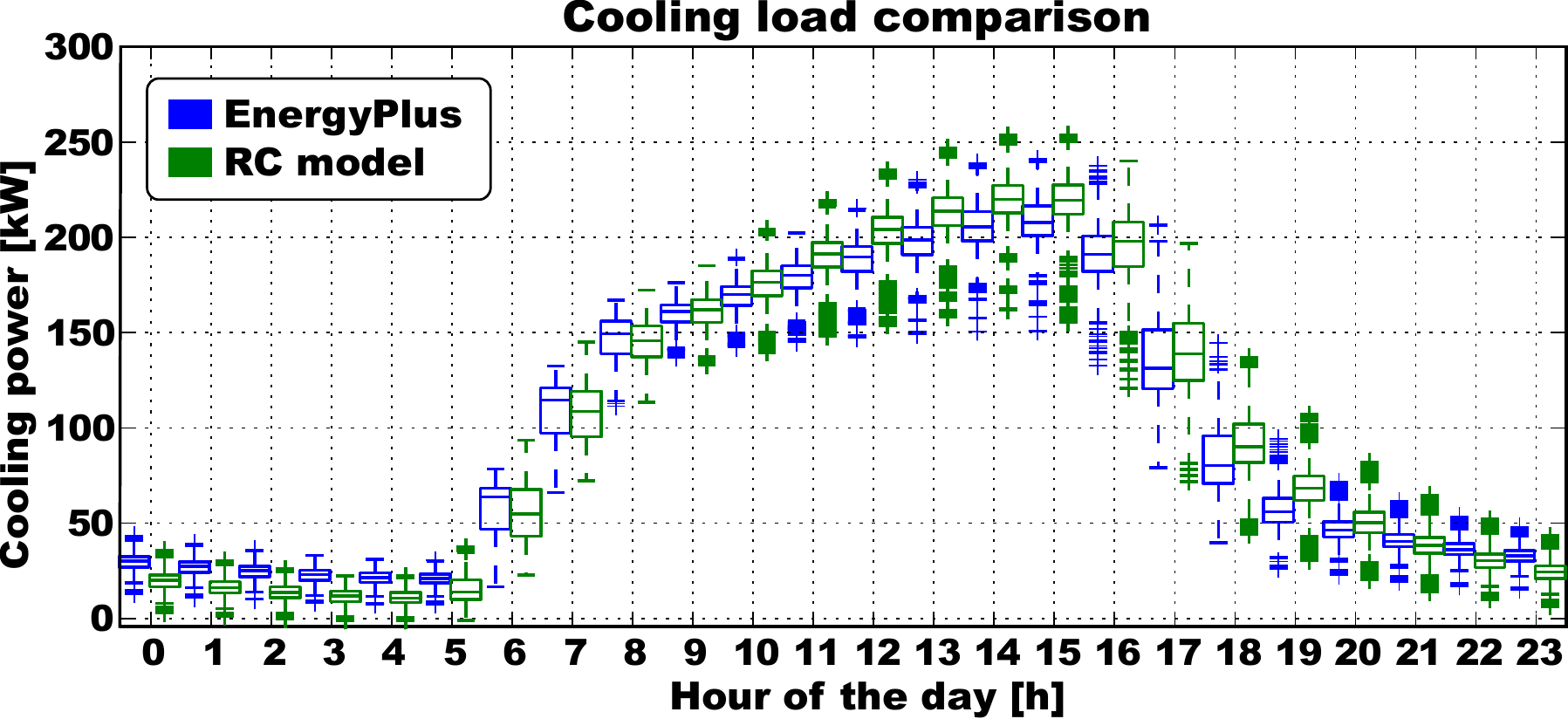}
\caption{Comparison of the cooling load predicted by \ep/ and the RC model.}
\label{fig:building_boxplot_comp}
\end{figure}

Figure (\ref{fig:building_scatterplot_comp}) shows a comparison across the hour of the day between the cooling load predicted by 
\ep/ and the RC model. All the simulation points for every hour of the day over the simulation period have been collected
and their distribution have been compared. The blue boxes shows the cooling load distribution for the \ep/ model while the
green boxes shows the distribution of the cooling load predicted using the RC model.
The highest relative errors evidenced in Figure (\ref{fig:building_scatterplot_comp}) happen exclusively at night
time when the nonlinearities of the long wave radiation exchange that are not captured by the RC model dominate
the heat transfer of the building. However Figure (\ref{fig:building_scatterplot_comp}) help us to put in perspective this error,
confirming that even if it's relative high it's impact on the ability of the RC mdoel to predict the cooling load can be neglected.

Under these assumptions it's possible to consider the RC model detailed enough to describe the thermal dynamics of the
building to be able to predict with a good accuracy its cooling load. The original RC model generated using the BRCM toolbox
had 106 state variables. The model has been reduced to a model that is able to predict the same averaged building temperature as
described in section \secRef{sec:Buildings_FMU}. The obtained reduced order model had 8 state variables and its difference between the full RC model is small enough (less than 1\% in relative terms) to be neglected for the sake of this study.

\subsection{Overview of the Use Cases}
\begin{figure}[htb]
    \centering
    \includegraphics[width = 0.75\columnwidth]{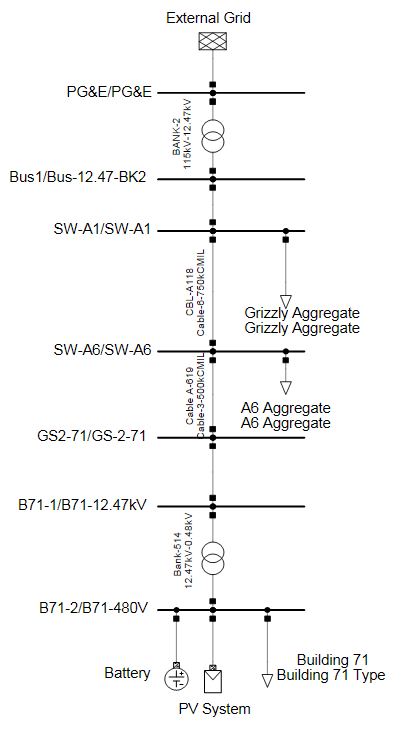}
    \caption{\pFactory/ model of the LBNL distribution system to Building 71.
    The rest of the LBNL, except for the Building 71, are modelled as aggregate loads on buses SW-A1 and SW-A6.}\label{fig:B71_Powerfactory}
\end{figure}

\begin{figure}[htb] 
\centering 
\def\svgwidth{1\linewidth}  
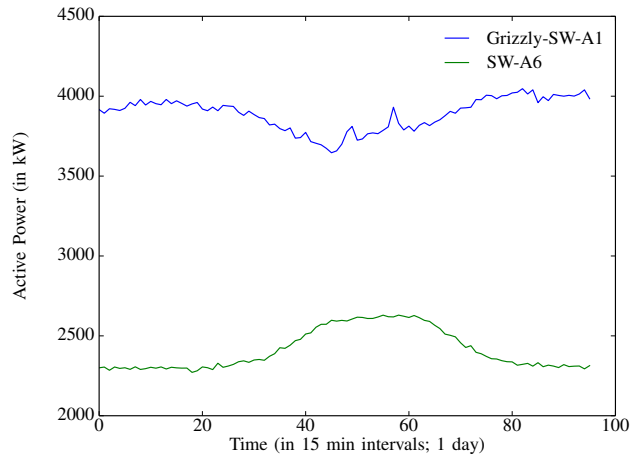
\caption{Active Power Consumption of the Aggregate Loads.} 
\label{fig:GrizzlyPandA6P} 
\end{figure}

Both cases use the LBNL distribution network and Building 71. As shown in Fig.~\ref{fig:B71_Powerfactory}, the LBNL distribution network represents the path from the point of common coupling with PG\&E, down to Building 71.
The remainder of the distribution system loads are modeled as aggregated loads connected to two swithcing substations along this path. Real 15-minute data
were used for the two aggregate loads.
For Bus SW-A1, real reactive power was used, while for SW-A6, a power factor of 0.94 inductive was assumed. The active power consumption of the aggregate loads in SW-A1 and SW-A6 are shown in Fig.~\ref{fig:GrizzlyPandA6P}.

\begin{figure}[htb] 
\centering 
\def\svgwidth{1\linewidth}  
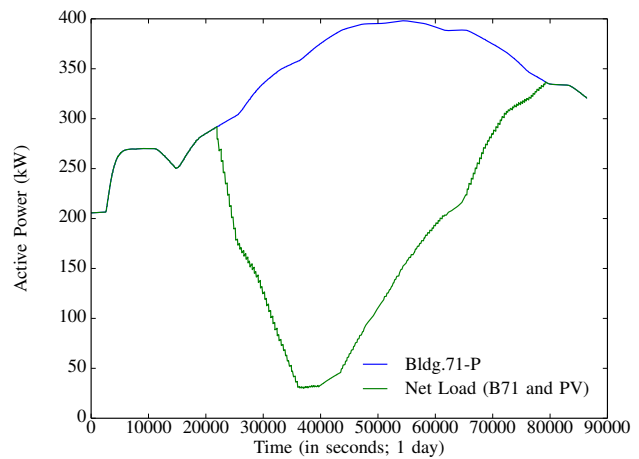
\caption{Active Power Consumption of Building 71 without Demand Response (blue) and Net Load at Bus B71 (green; sum of load and PV infeed).} 
\label{fig:B71_P_Q_NetLoadminusPV} 
\end{figure}

The modeling of Building 71 has been detailed in Section~\ref{sec:building_calibration}. To study the interaction of loads during high penetration of DER, we have assumed a solar PV plant of 340 kWp and a battery connected at the same bus. 
The active power consumption of Building 71 and the net load  (Building 71 and solar PV) demanded at Bus B71 is shown in Fig.~\ref{fig:B71_P_Q_NetLoadminusPV}. 
As we did not have available data for the reactive power consumption of Building 71, we assumed a constant power factor of 0.96 inductive.

We present two use cases in the following sections to demonstrate the capabilities of VirGIL. The first applies demand response actions in Building 71 to reduce the cable and transformer loading. The second applies Volt-Var control so that the voltage at Bus B71 follow specified setpoints.


\subsection{Demand Response in Building 71}
\label{sec:case_LBNL_DR}

\begin{figure}[htb] 
\centering 
\def\svgwidth{1\linewidth}  
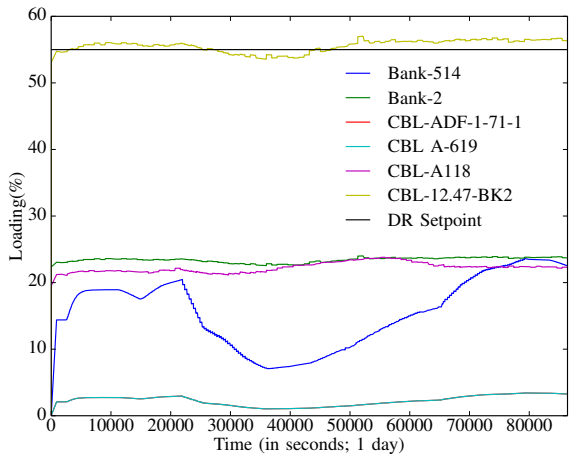
\caption{Cable and Transformer Loading (no Demand Response).} 
\label{fig:LineTrafoLoading_NoShed} 
\end{figure}

Figure~\ref{fig:LineTrafoLoading_NoShed} presents the transformer and cable loadings of the LBNL network during a typical day, when no demand response actions are taken. 
As a secure and reliable power supply is central for the operation of a National Laboratory and the experiments that are taking place in it, we observe that the maximum loading of all cables and transformers does not exceed 60\%. 

In order to demonstrate the capabilities of VirGIL, we set the threshold for demand response actions at 55\%. 
If any cable or transformer loading exceeds the 55\% threshold, an automated demand response signal is sent to the building, reducing its power consumption by 20\%. 
The Communications FMU ensures that the communication between the controller and the Building follows the OpenADR standard. Besides plug loads and lighting, which are considered constant during the day in this case, the main building load is HVAC cooling. 
As soon as the Building receives the DR signal, it has lookup tables that transform the power reduction to increased setpoints for the HVAC operation, as higher operating temperatures reduce the necessary cooling power.

\begin{figure}[htb] 
\centering 
\def\svgwidth{1\linewidth}  
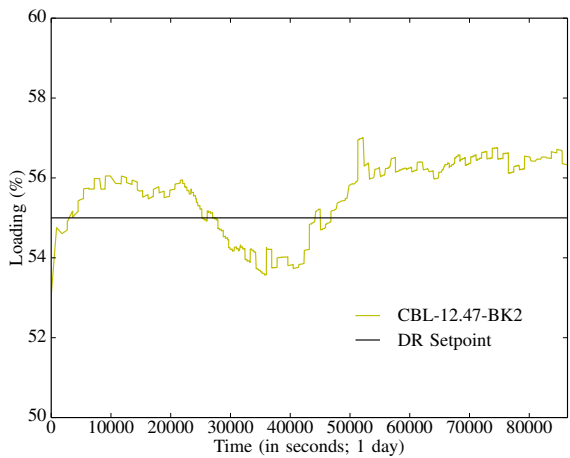
\caption{Loading of the most critical cable and setpoint for Demand Response actions.} 
\label{fig:LineTrafoLoading_onlycritical_NoShed} 
\end{figure}

Figure~\ref{fig:LineTrafoLoading_onlycritical_NoShed} zooms in the most critical cable loading, when no DR action is taken. We expect DR signals to be issued from midnight till 9am in the morning, and then again after 1pm in the afternoon.

\begin{figure}[htb] 
\centering 
\def\svgwidth{1\linewidth}  
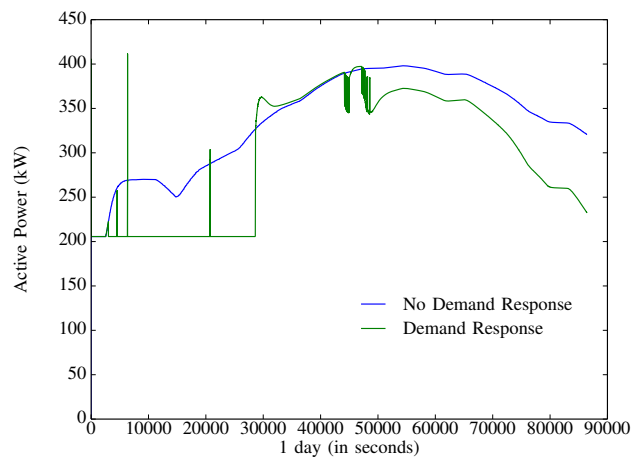
\caption{Active power consumption of Building 71 with and without Demand Response Actions.} 
\label{fig:B71_P_Q_Shed_NoShed} 
\end{figure}

\begin{figure}[htb] 
\centering 
\def\svgwidth{1\linewidth}  
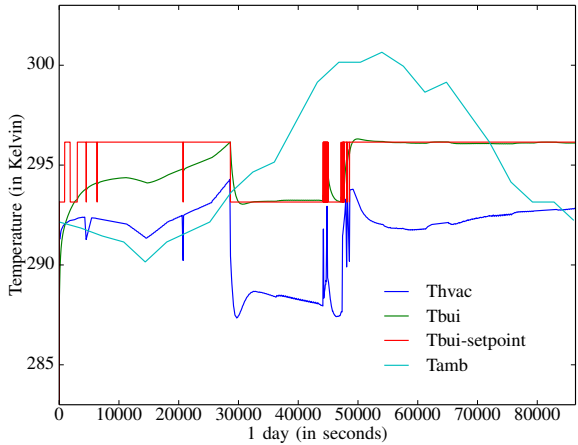
\caption{Temperature variations in Building 71 with active Demand Response. Tamb: ambient temperature; Tbui-setpoint: setpoint of the temperature setpoint inside the building (varies based on demand response actions); Tbui: average temperature inside the building; Thvac: setpoint at the HVAC controller.} 
\label{fig:TambThvacTbuiSP_Shed_delay30sec} 
\end{figure}

Figure~\ref{fig:B71_P_Q_Shed_NoShed} shows the active power consumption with DR and how this differs from the baseline for that day. 
To better understand this figure, we should first examine the sequence of events inside the building. Figure~\ref{fig:TambThvacTbuiSP_Shed_delay30sec} presents the temperature variations in Building 71 for this day. ``Tamb'' stands for the ambient temperature. 
``Tbui-setpoint'' is the setpoint for the average zone temperature inside the building. In case a DR event takes place, this setpoint changes. 
In our case, as we have cooling load, the setpoint increases.
``Tbui'' represents the actual average zone temperature of the building. Ideally, ``Tbui'' should follow ``Tbui-setpoint''. 
``Thvac'' represents the temperature that is given as input to the HVAC system. This results from the actions of the PID controller which monitors the building temperature and automatically adjusts the HVAC setpoints.
As expected, we see that the ``Tbui-setpoint'' increases during the first 9 hours of the day, then decreases to its nominal temperature of 293 Kelvin ($20^{o} C$) and increases back again after 1pm. 
We observe some oscillations in the DR signal at t=45000 s. This is because the cable loading is marginally above or below 55\%. 
As the day starts and the ambient temperature increases, we see that ``Tbui'' increases as well. at about 9am (t=30'000 s) it reaches the ``Tbui-setpoint''. From that point on the average zone temperature follows the ``Tbui-setpoint''. 
Going back to Fig.~\ref{fig:B71_P_Q_Shed_NoShed}, we observe that if DR is activated, the building consumes no additional power till about 9am, when the DR signal ends. If the ``Tbui-setpoint'' remained at 293 K, then Building 71 would have increased it socnsumption already from about t=2600 s. Similarly, we see that after about 1 pm (t=50'000 s), when DR is activated again, the building consumption is reduced in comparison to the baseline.

\begin{figure}[htb] 
\centering 
\def\svgwidth{1\linewidth}  
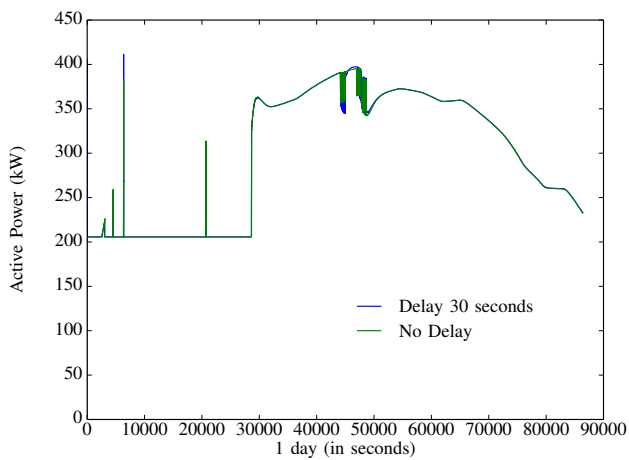
\caption{Active power consumption of Building 71 with demand response, with and without signal communication delays. }
\label{fig:B71_P_Q_Shed_Delay_NoDelay}  
\end{figure}

Figure~\ref{fig:B71_P_Q_Shed_Delay_NoDelay} compares the active power consumption of Building 71 when there are no communication delays. 
In our base case, shown in Fig.~\ref{fig:B71_P_Q_Shed_NoShed}, the polling frequency of the DR client in the Communications FMU was set at 30 s. This means that every DR signal was sent with 30~s. The actual communication delays that were also modelled in this setup did not exceed on average  ~500 ms.
In Fig.~\ref{fig:B71_P_Q_Shed_Delay_NoDelay}, we do not observe major differences between the two cases. This is probably due to the fact that the building dynamics are slow enough to not be significantly affected by a 30~s delay. Still, We observe that at about t=45'000~s, the oscillations of the active power have a higher magnitude when the singal is transmitted with no delay. This is expected due to the more direct response to the signal.

Concluding this use case, we see how VirGIL is able to accurately model and simulate the interactions between buildings, communication, and power systems. 
We have observed how a power system event leads to a DR signal, and how this affects the building operation. 
At the same time, we were able to represent the effect of the communication delays, and measure the effect of the building actions back to the power grid.

\subsection{Volt-Var Control at Bus B71}
\label{sec:case_LBNL_VoltVar}

In this use case, we demonstrate how VirGIL can be used for Volt-Var control. In the LBNL network we have installed three micro-phasor measurement units ($\mu$PMUs) \cite{Stewart_uPMU}. One of them is located at Bus SW-A6 and one more at Bus B71. $\mu$PMUs are units that can measure with high fidelity voltage, current, and voltage angle.

In this case we assume that we receive as inputs the voltage from the $\mu$PMU measurements at Buses SW-A6 and B71. The goal is that the voltage at Bus B71 should follow the voltage at SW-A6, by appropriately controlling the reactive power infeed of the battery.

The controller solves the following equation in order to find the necessary reactive power infeed:

\begin{equation}
{U_1}^2=(U_{2}+(R\cdot P+X \cdot Q)/U_{2})^2+(X \cdot P-R \cdot Q)^2/(U_{2})^2
\end{equation}

where:
\begin{align}
U_1&=U_{\textnormal{SW-A6}}\\
U_2&=U_{\textnormal{B71}}\\
P&=P_{B71}-P_{BAT}-P_{PV}\\
Q&=P_{B71}-Q_{BAT}-Q_{PV}\\
R&=R_{\textnormal{Bank-514}}+R_{\textnormal{CBL-ADF-1-71-1}}+R_{\textnormal{A-619}}\\
X&=R_{\textnormal{Bank-514}}+X_{\textnormal{CBL-ADF-1-71-1}}+X_{\textnormal{A-619}}
\end{align}

\begin{figure}[htb] 
\centering 
\def\svgwidth{1\linewidth}  
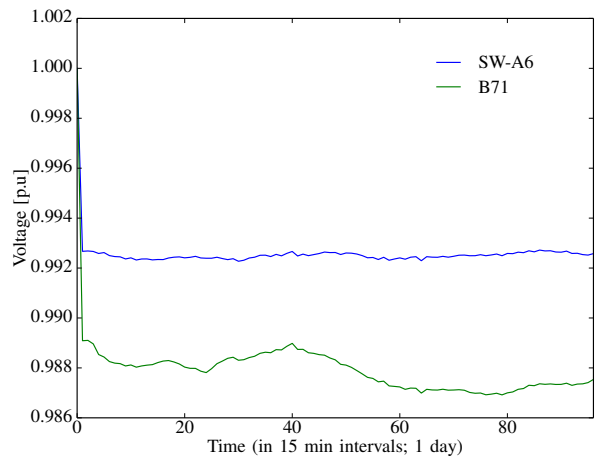
\caption{Without Volt-Var Control: Voltage level of Bus SW-A6 and Bus B71.}
\label{fig:V_A6_V_B71_noBatP_noControl} 
\end{figure}

Figure~\ref{fig:V_A6_V_B71_noBatP_noControl} presents the voltage at the two buses if no volt-var control actions take place. We can observe how the PV infeed, starting at about 7am, increases the voltage momentarily, while in general the voltage level at Bus B71 is decreasing as the building consumption increases. Once again, we observe that the LBNL network is sufficiently (over)dimensioned so that we do not observe significant voltage drops at the end of the feeders. Still this use case demonstrates VirGIL performance and characteristics. 

\begin{figure}[htb] 
\centering 
\def\svgwidth{1\linewidth}  
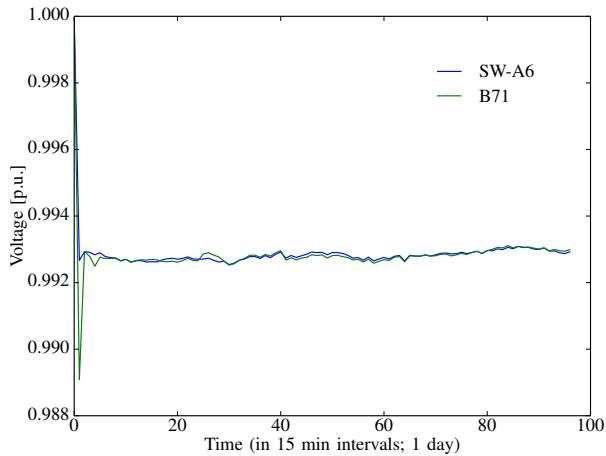
\caption{With Volt-Var Control: Voltage level of Bus SW-A6 and Bus B71.}
\label{fig:V_A6_V_B71_noBatP_withControl} 
\end{figure}

\begin{figure}[htb] 
\centering 
\def\svgwidth{1\linewidth}  
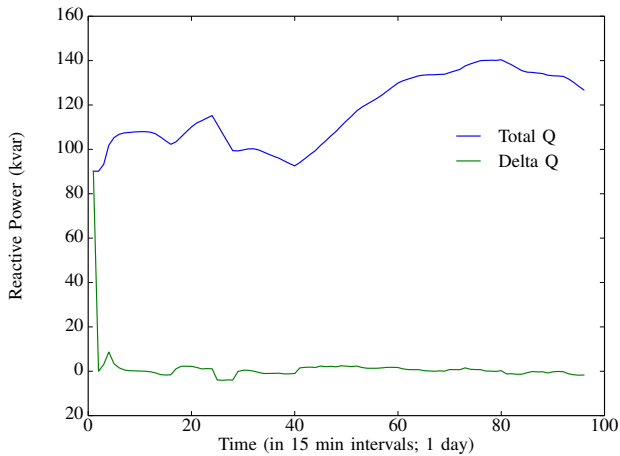
\caption{Reactive power injection of the battery. Total Q corresponds to the reactive power injected. Delta Q is the change in reactive power from the previous timestep.} 
\label{fig:Qtotalnecessary_DQ_SP_noBatP_withControl} 
\end{figure}

Figure~\ref{fig:V_A6_V_B71_noBatP_withControl} presents the same voltages, but with volt-var control, so that $V_{B71}$ to track the voltage $V_{SW-A6}$ at Bus SW-A6. The required reactive power infeed from the battery is presented in Fig.~\ref{fig:Qtotalnecessary_DQ_SP_noBatP_withControl}. In the same figure, we also present the $\Delta Q$, i.e. the difference by which the $Q$ setpoint should be adjusted from one timestep to another. 

\section{Conclusions}
\label{sec:conclusions}

The Virtual Grid Integration Laboratory (VirGIL) creates a modular co-simulation platform for studying in detail the impact of demand response and other controls on power systems.
The platform coordinates commercial software such as \pFactory/, open-source packages such as the Modelica Buildings Library, communication simulation tools such as OMNET++ and bespoke models (such as the Buildings FMU described above).
Using commercial and trusted power system software is expected to lower the barriers for adoption of simulation and optimization tools by power system operators, allowing them to test, improve, and deploy new practices, e.g., efficiently integrating demand response in their daily operation.
VirGIL uses the industry-standard \fmInterface/ (FMI), which encourages a modular approach to instantiating and sharing models.


This paper presented the development of FMUs for Co-Simulation for Power Systems, Communications, and Control, and one FMU for Model Exchange for Building modeling and control. Real network and consumption data were used as parameters and inputs to these FMUs to model part of the LBNL distribution grid, and to couple the grid to a reduced-order physics-based model of a real building that implements a simple demand response protocol. A full representation of the communications network was also included.
To our knowledge, this is the first time that a co-simulation platform couples commercial power system software such as Powerfactory with building and communications models to study the impact of demand response actions on the distributions grid. A further contribution of this paper is the full integration of the Quantized State System (QSS) methods for simulation in VirGIL.

\pt{II}, the VirGIL implementation framework, handles both continuous and discrete-event simulation, and supports both FMI for model exchange and co-simulation.

Future extensions of this work include electric vehicles, power system optimization, and advanced building controls.
Use cases will include demand response applications for volt/var optimization, 3-phase asymmetries, and distribution system planning.
Real case studies, such as the demand response potential and impact in the region of the decommissioned San Onofre Nuclear Generating Station (SONGS), will be simulated in VirGIL and presented.


\section*{Acknowledgements}
We thank Edward Lee and Christopher Brooks from UC Berkeley for their support in the FMI and QSS implementation in Ptolemy II. This work was supported by Laboratory Directed Research and Development (LDRD) funding from Berkeley Lab, provided by the Director, Office of Science, of the U.S. Department of Energy under Contract No. DE-AC02-05CH11231.

\bibliographystyle{IEEEtran}
\bibliography{includes/IEEE_Proceedings_biblio}
\vspace{-0.3cm}


\end{document}